# The Episodic Time Interpretation of Reality

### *or, Making the World Safe for Quantum Mechanics*

**Daniel S. Klotzer[1]**

[1] Washington University, Dept. of Physics - Compton Hall, 1 Brookings Drive - Campus Box 1105, St. Louis, Mo., USA 63130; e-mail: klotzer@physics.wustl.edu

## Abstract

A notably enhanced comprehension of the underlying meaning of quantum observations is achieved via a novel premise.  Assessments, from first principles, are made of unexamined presumptions that lie at the heart of both conventional conceptions of the nature of physical existence and most interpretations of Quantum Mechanics.  An alternative hypothesis, termed Episodic Time Inhabitation, is proposed to resolve major Quantum quandaries, including the EPR paradox.  A logical argument gedanken experiment demonstrates that the conventional presumptions appear to create causal loops, and that Episodic Time Inhabitation avoids them.  A physically realizable version of the gedanken experiment is outlined, and currently testable predictions for its outcome are made.  Consequences of the experiment are summarized and ramifications of the hypothesis are discussed.

## BACKGROUND

Throughout the often contentious history of Quantum Mechanics (QM), resolving the intrinsic nature of physical reality has been at the crux of many disputes. The customary approach used to examine these issues appears misdirected, however. The conventional supposition is that the formulation of QM must correspond with what physical reality "should" be. In fact, how physical reality is conceived of should be brought into correspondence with QM. Of all presently testable physical theories, QM is arguably the most accurate. Since QM provides the most certain indication we currently have of the actual nature of physical reality, it is more reasonable to alter what is less certain: our conceptions of physical reality.

The Copenhagen Interpretation (Bohr, N., 1935) asserts that physical reality consists of only what can be observed. The existence of a reality outside of an observation is not only undefined, but is considered to not even exist, prior to the observation. Due in large part to its proponents' vociferous ardor, the Copenhagen Interpretation became the de-facto standard, to the point that even among many physicists there is only limited awareness that it is an opinion, not fact. It is, of course, possible to conjure all forms of gedanken scenarios that illustrate the apparent irrationality of this conceit, such as Schroedinger's famous cat (Schrödinger, E., 1935). A more pointed questioning of the completeness of the Copenhagen Interpretation was explicated in Einstein, Podolsky, and Rosen's 1935 paper (Einstein, A., Podolsky, B., and Rosen, N., 1935) describing the counterintuitive consequences of entanglement that came to be known as the EPR paradox. A more detailed scrutiny of the Copenhagen interpretation's validity (as well as the veracity of other interpretations) follows in a later discussion of a proposed experiment.

Among the alternatives to the Copenhagen Interpretation are interpretations that invoke "hidden variables". These theories suppose that ***definite*** values do exist for those quantities that are not definitively knowable in conventional QM, but that these values are not observable (as opposed to the Copenhagen Interpretation which postulates that these quantities do not have definite values prior to their observation). Bohmian Mechanics (Bohm, D., 1952) is among the foremost of these hidden variable theories. Bohm hypothesized a "pilot wave", essentially consisting of the influences of everything else not being observed, which governs quantum phenomena though it can not – by definition – be observed. Hidden variable theories all have the "strength" of defining phenomena that is not observable, and hence are definitions which seemingly cannot be disproved. Of course, proving them valid is also a problem. Hidden variable theories are born of faith in a particular conception of physical reality, a philosophical attitude akin to naïve realism.

Bell's seminal work in 1964, "On the Einstein-Podolsky-Rosen paradox" (Bell, J. S., 1964) carefully tested the suppositions of hidden variable theories and the general postulates of a naïve realist perspective against the well-known quantum entanglement phenomena which is generally termed "action-at-a-distance". The resulting Bell Inequalities demonstrate that these suppositions and postulates (which are often lumped together as the "classical" perspective) predict results that are manifestly divergent from the results predicted by the "quantum" perspective. Experimental results (Freedman, S. J. and Clauser, J. F., 1972; Aspect, A., 1999; Grangier, P., 2001) have thus far agreed with the predictions of the quantum perspective with sufficient strength to appear to settle the issue. It has most commonly been assumed that the issue settled is whether or not hidden variable theories have any credence, though the results only show that a combination of these theories' suppositions – the naïve realist postulates of the "classical" perspective – and Bell's assumptions (Bell, J. S., 1964) are not supported.

Among the core postulates of the naïve realist perspective, and a fundamental underpinning of a substantial proportion of investigators' attitudes, is the principal of local action. The gist of the EPR paradox is that "action-at-a-distance", unconstrained by the proscriptions of special relativity, appears irreconcilable with what the nature of physical reality is believed to be. Curiously though, efforts to define localization, from first principles or even empirically, are practically non-existent. Localization is almost always addressed only by its absence. It is often useful to work in an environment where two sets of experimental apparatuses have, in the vernacular of relativity, a space-like separation. These apparatuses are presumed, according to special relativity, to be unable to interact within the experimental time-span. Such conditions are specifically employed to explore what are defined as non-local phenomena. However, exactly what defines "local" and where or how the transition from non-local to local occurs is seldom addressed. A new perspective on this issue is the genesis of the present hypothesis.

An inspection of how Bell addressed the issue of localization is instructive. In the Introduction (Bell, J. S., 1964) Bell addresses what he relates as the argument of the EPR authors in favor of supplemental, hidden variables to "restore to the theory (of QM) causality and locality". Bell then continues: "It is the requirement of locality, or more precisely that the result of a measurement on one system be unaffected by operations on a distant system with which it has interacted in the past,…." In a more explicit detailing from the Formulation section (Bell, J. S., 1964), referring to measurements of an entangled pair of spin ½ particles in an example from Bohm and Aharonov (Bohm, D. and Aharonov, Y., 1957) : "…we make the hypothesis,…, that if the two measurements are made at places remote from one another the orientation of one magnet does not influence the result obtained with the other. Since we can predict in advance the result of measuring any chosen component of $\sigma_2$, by previously measuring the same component of $\sigma_1$, it follows that the result of any such measurement must be predetermined." What is revealing is that while Bell carefully considered the assumptions he chose to adopt (as did Born, Einstein, Podolsky, Rosen, and Bohm), there are also assumptions he, and others, make prior to their careful considerations. These assumptions are virtually inherent to all examiners' customary states of mind, and are preliminary to any measured evaluations. They will be referred to here as "presumptions" – perhaps not unconscious, but apparently unconsidered and unquestioned.

Of greatest relevance here is the presumption that localization refers to spatial dimensions only. When referring to "locality" Bell exclusively considers "distant systems" and "places remote from one another". Regardless of the number of additional dimensions that may eventually be determined to comprise physical reality, there is no question that there are at least four: three spatial dimensions and time. Contemplating locations in space, with the advent of QM, has become a complicated process. Beyond just finding the result of the determination of a physical entity's location, QM also requires accounting for the measurement process as well as the uncertainty (and hence the "spread") of the entity about the determined location. It now is effectively second nature to conceive of physical entities as occupying semi-nebulous "footprints" in space, and physical entities that possess definite locations or are point particles are more difficult to imagine. Customarily though, the same degree of scrutiny is not applied to locations in time.

A normally overlooked aspect of physical existence in time is decisive: the nature of "how" physical entities exist in the temporal dimension, which is termed here the "inhabitation" of time. The nearly universal presumption is that time is physically inhabited in an ***ephemeral*** present moment. In plain language, "now" is presumed to be vanishingly brief. The question of how time is occupied physically is simply not addressed even by those that focus on examining our fundamental tenets of the nature of reality. For example, Mermin (Mermin, N. D., 1998) states that: "Physics has nothing to do with such notions. It knows nothing of *now* and deals only with correlations between one time and another." Ephemeral Time Inhabitation (Ephemeral TI) is fairly well exemplified by a delta function, $\delta(t)$ with argument $t \equiv$ the present moment. When calculating, the "footprint" in time of the ephemeral present moment is mathematically expressed by the infinitesimally brief integration interval $dt$. Clearly, ephemeral is a fitting appellation. A related, and important, note is that the characteristics of the integration interval $dt$ are primarily due to the requirements of the mathematical basis of calculus. These characteristics have proven of spectacular utility in classical and even quantum mechanical calculations. But, do physics principles command that they are the only possible way to describe the nature of time inhabitation? And if these principles don't, are there any reasonable alternatives, or motivating factors for imagining that alternatives might exist?

Inspiration for the alternative that is the progenitor of the present hypothesis arises from contemplation of the EPR paradox. Ample experimental evidence (Aspect, A., 1999; Grangier, P., 2001; Tittel, W., Brendel, J., Zbinden, H. and Gisin, N., 1998) has shown that the EPR effect is real. Special relativity, to all indications, is also valid. How, then, can the "action-at-a-distance" occur across divides which are evidently unbridgeable? Even if one embraces the precepts of the Copenhagen or many-worlds interpretations, the question is not resolved by being couched in terminologies such as "collapse of the wave function". Later, it will be seen that philosophical arguments are unnecessary to confound these interpretations. A straightforward, and eminently achievable, gedanken experiment is proposed which impugns their validity. In appendix B means for physically realizing a critical part of this experiment with currently accessible technology is outlined.

Various explanations have been proposed for questions raised by the EPR paradox. One species of conjectures speculate that a form of "back-action" moves backward through time. These theories envision that, since the physical entity under consideration had not been observed prior to the entanglement ending measurement, the entity could conceivably be affected by future events without confounding any *known* history. If there were to be such a change in a known history, the occurrence of a casual loop – thought by most to be a highly undesirable event – would be possible (as will be seen in the later explication of the proposed experiment).

A prominent theory of this type is Cramer's Transactional Interpretation of Quantum Mechanics (Cramer, J. G., 1986) wherein Cramer states that "The basic element of the transactional interpretation is an emitter-absorber transaction through the exchange of advanced and retarded waves". Essentially, the Transactional Interpretation "cleans" up the mess left by QM's action-at-a-distance conflict with relativistic constraints by positing a "measurement" wave traveling backward in time. The measurement wave returns to the point of origin of the quantum entity which is "collapsed" by the measurement event, and thereby imparts the impact of the measurement event to the point of origin so that, for example, a quantum entity which is entangled with the measured entity will be able to receive the effects of the measurement event without an information transfer that contravenes relativistic constraints. The backward traveling wave is defined to be allowable because it only impacts the worldline of an, until then, unmeasured entity, and will thus not alter any "actual" part of physical reality. The Transactional Interpretation implicitly, and knowingly, incorporates the Observer-Actuated paradigm. It also embraces the presumption of Ephemeral TI, and will be seen to suffer the same difficulties as the other theories that are based on this presumption.

## EPISODIC TIME INHABITATION

### Introduction

The EPR paradox has crystallized the understanding that QM's conceptual problems are, at their core, a question of localization. But not solely a question of where the boundary between local and non-local lies. The core of the question is: In which dimensions must localization be examined? Clearly, localization in space is essential. But is that all? For the EPR effect to occur there must be some form of what appears to be an "instant" connection between space-like separated events – unless one subscribes to an observer-actuated interpretation (O-AI) such as the Copenhagen, Transactional, or Many-Worlds interpretations which will be discussed later. These separated events are connected only by entangled entities that diverged spatially well before the "instant" connection. The separation between these events is conventionally seen as a separation in space, though it can also be seen as a separation in time. Either the communication between space-like separated, entangled events contravenes relativistic speed limits on passage through space, or it contradicts standard assumptions about time-passage and causality.

At the heart of the present hypothesis is the assessment that localization in time is the paramount issue. The habitual presumption that the temporal dimension(s) is only inhabited ephemerally is discarded. Rather, it is proposed here that time is ***not*** physically inhabited in an infinitesimally brief present moment. Hence, a "present moment" has a duration of at least some finite extent. Moreover, it is further advanced that a present moment not only has a non-infinitesimal span, but that the interval of time a present moment encompasses is variable. The phrase "present moment" is used only grudgingly, dictated by the need to communicate in familiar terms. The common conception of a "moment" does not match the present conjecture. A new label for the period of time occupied by a physical entity's "present" is required. The "footprint" in time an entity inhabits is henceforth defined as "the present episode". Correspondingly, the present theory is identified as Episodic Time Inhabitation (Episodic TI). This theory is not termed an interpretation of QM, because it is actually an interpretation of physical reality. QM is a mathematical system for predicting the results of physical experiments. The results of QM measurements are the most accurate confirmations known of any theory's predictions about physical existence. When constructing an ontological interpretation, one's starting point should be from the area of greatest certainty. Regardless of comfort level, that area of greatest certainty is QM. Evidently, what needs to be (re)interpreted are the prevailing conceptions of reality.

In approaching Episodic TI's reinterpretation of reality the limitations of common language, and hence normal modes of thought, must be recognized. For example, in the third sentence of the immediately preceding paragraph, the term "duration" is used in reference to the length of the present moment. This term can be misleading, since it is commonly used to mean the length of a passage *through* time. Here though, in referring to the extent of a present episode, duration means the span of time that is contained within the present episode. Episodic TI can be envisioned as the occupation of a length of the temporal dimension, similar to the occupation of spatial dimensions. Hence the use of the term "footprint in time".

The term "simultaneous" should also be used with care. At first glance, it seems natural to describe the inhabitation of a present episode as being "the simultaneous occurrence of all events within that episode". This description intermixes two related, but distinct, attributes of physical existence in time. Time has both a meter and an ordering. These are elements of the composition of the time dimension, a constituent of spacetime. Whether the structure of spacetime is eventually found to be discrete or continuous, singly or multiply dimensioned, or some other composition and organization (including realms in which spacetime may undergo a fundamental phase transition such as intervals less than the Planck length or time), these qualities are characteristics of the temporal degree of freedom. A second attribute, which is the focus of the present theory, is how physical entities exist within, i.e. occupy, the temporal degree(s) of freedom.

Within a given specification of time's structure, the form of inhabitation of that time structure can conceivably vary. Moreover, the manner of time's inhabitation can conceivably be similar across differing compositions of time. The ordinary understanding of a concept like "simultaneous" implicitly employs the precepts of Ephemeral TI. In Ephemeral Time Inhabitation, different locations in time are presumed to be different present moments. Special relativity shows that no absolute ordering is definable between different reference frames.  For Ephemeral TI, distinctly separate points in time within a single reference frame cannot be concurrently occupied by a physical entity.  For Episodic TI, they can, and are.

### **Indications**

At minimum, Episodic TI is a marked departure from accepted wisdom.  The first priority is to discover if this divergence is suggested or warranted, much less affirmed, by the available evidence.  The conceptual inclinations that have led to the formulation of Episodic TI are an affinity for causal relationships enacted by a relativistically sensible local interaction and a distaste for an observer-actuated reality.  The only local interaction connecting a pair of EPR particles occurs well before their measurement, which suggests that a connection bridging the boundary between past and present is needed.  The leap to Episodic TI was prompted by the lack of satisfactory explications [such as the "Special Lorentz Frames" of Hardy's PR6 and PR7 (Hardy, L., 2001; Percival, I. C., 2000) which attempt, unsuccessfully, to define simultaneity for events with spacelike separation] for the relativistic quandaries raised by the EPR paradox.  If the boundary between past and present is traversable, or even erasable, a greater understanding might ensue.

Among the most perilous conditions which can afflict any theory of physical existence is an allowance of causal loops.  A coupling between what is defined as "past" and "present" within a given reference frame could create causal loops, as discussed in detail later.  Ephemeral TI, when applied to a proposed Causal Loop Creation Experiment, generates such a coupling.  One option for avoiding this coupling is to effectively broaden the meaning of what is called the "present".  The advisability of this option can be illuminated by assessing its benefits versus its costs.  Turning first to the potential costs, there is no question of the immense utility and success garnered from modeling physical existence with Ephemeral TI.  All of calculus and hence a vast amount of mathematical physics is incumbent on the *dt* paradigm of time inhabitation.  This model is constructed upon a theory of approximations taken to a limit.  It is not constructed on physics first principles.  The limitations on solvable problems, due to calculus' requirements of differentiable functions and other constraints, fail to include a multitude of actual physical phenomena.  The proscriptions of a mathematical model, no matter how customary or fruitful, should not prevent the use of other potentially valuable models. On the other hand, the wealth of understanding provided by Ephemeral TI should not be squandered lightly. Ephemeral TI is an important perspective for analyzing physical existence. It would behoove any new model, and Episodic TI is no exception, to avoid relinquishing the accomplishments of Ephemeral TI, if at all possible. Fortuitously, resolving this concern provides substantial new insights. Crossing between Ephemeral TI and Episodic TI perspectives is comparable to shifting between classical and quantum mechanics. In fact, it is directly analogous. The classical perspective corresponds to Ephemeral TI, and the quantum perspective corresponds to Episodic TI. The seeming incongruity of quantum phenomena is a direct result of observing Episodic TI phenomena from an Ephemeral TI perspective. Grappling with a potential problem of Episodic TI has led to a considerable dividend.

A persisting difficulty interwoven with the other questions regarding QM is the issue of the "boundary" between the classical and the quantum realms. Quantum phenomena were originally thought to be the province of a greatly constrained, though important domain. These constraints are no longer immobile, or even very restrictive. With virtually every succeeding month, new evidence of mesoscopic quantum phenomena is documented (Brezger, B., Hackermüller, L., Uttenthaler, S., Petschinka, J., Arndt, M. and Zeilinger, A., 2002; Friedman, J. R., Patel, V., Chen, W., Tolpygo, S. K. and Lukens, J. E., 2000). Most explanations for the existence of the boundary and its variability invoke causes related to the nature of the investigatory method, such as a superposition of the (usually) microscopic quantum phenomena and the macroscopic measurement apparatus (von Neumann, J., 1932); explore decohering mechanisms (Myatt, C. J., King, B. E., Turchette, Q. A., Sackett, C. A., Kielpinski, D., Itano, W. M., Monroe, C. and Wineland, D. J., 2000); and/or posit that classical reality arises as a statistical artifact of the probabilistic formalism of orthodox QM (von Neumann, J., 1932).

It is suggested that a comparison of Ephemeral TI and Episodic TI perspectives will yield fundamental insights into the relation between quantum and classical phenomena. The vital distinction between the quantum and classical realms is a product of the size of a "present episode". As the span of an episode shortens, the behavior of a physical entity more nearly approximates the classical. In the limit of an infinitesimally brief episode, the physical entity manifests Ephemeral TI and thus behaves classically. It is proposed that the transition between classical and quantum behavior is directly related to the length of the "present episode". The classical character of common day experiences is thus due to the exceptionally high frequency of interactions among the dense aggregations of individual entities that comprise normal matter. In ordinary matter the high density of entities, such as atoms, produce virtually constant interactions. The rate of interaction may even be so high that the duration of a particular interaction is considerably greater than the time span between interactions, in which case it may not even be sensible to discuss individual interactions, since they will be overlapping and essentially incessant. The typical scale of episode length would then approach the order of the $dt$ limit, and the classical approximation would be highly accurate for matter as a whole. From an estimate of an episode length, a judgment of the time-scale of a phenomena under consideration, and a relation to the time-span specifications of the observing mechanism, the result of an observation of a system can be ascribed a relative position between Episodic TI (quantum) and Ephemeral TI (classical) behavior. Rather than a statistical averaging across innumerable individual systems producing classical behavior in the aggregate, Episodic TI provides a discernable basis from first principles for assessing a system's degree of quantum vs. classical behavior, regardless of the scale, or quantity of systems under consideration.

An entity's "present episode" extends between an initiating and a concluding interaction. The interactions are labeled according to their order on the meter of time, not to reflect an order of the entity's experience. The initiating and concluding interactions are essentially boundary conditions on the status of the entity that inhabits the episode. Only at the boundaries of the episode can any information about the episode be measured, since said measurements are by definition interactions. A further propitious effect of Episodic TI is an enlightened comprehension of particle physics' arrow-of-time puzzle. In particle physics virtually all prevailing interaction models are symmetric with respect to time-reversal. Yet, events are not observed to flow backward in time, which is usually attributed to entropy considerations. Episodic TI furnishes an uncontrived integration of these disparate properties. All the time coordinates contained within a "present episode" are equivalently *now*. Any rationale involving a progression through these points, in whatever direction, is meaningless since their order is interchangeable. The boundary points are disparate though. The sum of events that constitutes the physical environment at the concluding boundary is greater than the sum of events that did, or could have occurred, at the initiating boundary. The physical environment at the concluding boundary includes events that ensued from the initiating interaction. Consequently, it is impossible for the environment of an episode's initiating interaction to replicate that of its concluding interaction. Hence, acquiescence to the arrow of time is inevitable outside of a "present episode", and irrelevant within one.

Exactly what happens within a "present episode" and how it relates to what is known of QM *is* relevant. The various "hidden variables" approaches were generally fabricated to reconcile naïve realism with quantum phenomena. In particular, these hidden variable formulations were intended to salvage a local reality from the inherent non-locality of O-AI's. Bell's inequalities are trumpeted by many as the demise of hidden variable theories. This is a misinterpretation, for Bell's work actually showed that if there were hidden variables, then physical existence must be spatially non-local. Although the experimental confirmations of Bell's inequalities clearly reveal that physical existence definitely is spatially non-local, they do not prove that hidden variable theories are necessarily false, and the present hypothesis is neither an exposition in favor of, nor against hidden variable theories. Since a chief impetus behind most hidden variable theories is to preserve spatial locality, these theories appear unlikely to succeed in fully achieving the goals of their originators. The debate over hidden variables is seen to be misguided, though, from an Episodic TI perspective. One could posit that the entirety of an episode, other than the part sampled in a measurement, is "hidden" to an observer that possesses an Ephemeral TI perspective, but this is only a matter of semantics. It is less than momentous whether or not an ephemeral observer can "see" all of an episode, or whether the unseen portion of an episode is defined as "actual" when it is not observed. These issues are side effects of the tension between O-AI's and a naïve realist perspective, while the real source of the problem is their mutual incorporation of Ephemeral TI's mischaracterization of physical existence.

Working with Episodic TI is problematic, since the development of human language and thought is rooted in Ephemeral TI. Auspiciously, the clarity and coherence of the Episodic TI interpretation improves as the range of quantum phenomena under study expands. Among the most striking of QM's defining characteristics is its probabilistic nature, for which Episodic TI provides a markedly more reasonable accounting than does Ephemeral TI. Principal obstacles to intuitive understandings of a probabilistic reality include the notions that an entity can be in a superposition of states, or have an indeterminate position; i.e., that something can be more than one thing, or in more than one place. The unstated, root source of these conceptual difficulties is that these nonsingular positions or states are occurring simultaneously (yet another unexamined presumption). For Ephemeral TI, simultaneous means "at the same point in time". With Episodic TI, simultaneous events (along a single worldline) can have different time coordinates. Standard (Ephemeral TI) approaches to physics inquiries are predicated on causal relationships that follow only timelike progressions, restrictions that Episodic TI does not share.

When working with Episodic TI, care in phrasing an inquiry is necessary. In physical terms, the "phrasing of an inquiry" might be an astronomical observation's chosen frequency range, or whether a QM experiment registers either an interference pattern or an individual particle's path. In the phrasing of questions about the statistical nature of QM measurements, the crux of the inquiry is whether or not there is a complete, definite specification (some of which may be hidden) of the state of an entity that predetermines the result of a measurement. As stated by Bell (Bell, J. S., 1964), in discussing an example advocated by Bohm and Aharonov (Bohm, D. and Aharonov, Y., 1957): "Since we can predict in advance the result of measuring any chosen component of $\sigma_2$, by previously measuring the same component of $\sigma_1$, it follows that the result of any such measurement must actually be predetermined." Bell presumes that events can occupy only a single point in time, and that any such point is separate from any other. Inevitably, the conclusions thus drawn are the product of these presumptions. The phrasing of the inquiry is not only critical for precisely understanding the physical parameters under investigation, but also for recognizing the assumptions (and presumptions) of the investigator.

Scrutiny of the nature of a position observation exemplifies inquiries into QM's probabilistic character. In general, the expectation values of an observable are determined by application of its operator to the relevant wave function (or density matrix) in a Hilbert space of appropriate dimensions. That the position operator **r** is relatively simple, and hence straightforward to discuss, does not restrict the generality of the discussion, because neither the form of the operator nor the nature of the wave function (or density matrix) is germane to the discussion. The inquiry is essentially: How can a physical entity not be at a specific place? Or rather, to make the implicit presumptions explicit: At one specific point in time, how can a physical entity not be at one specific place in space?

If Ephemeral TI describes physical existence, there is no reason to make the implicit presumptions explicit, since different points in time must be different present moments. For an Episodic TI description, a "present moment" is not synonymous with a "point in time", and the phrasing of the above inquiry is improper. Physical entities inhabit the time dimension, and the qualities of that inhabitation require an examination that is distinct from an examination of the time dimension, per se. The pertinent core of the above inquiry are the words "be at". An entity inhabits all the time coordinates of a present episode concurrently. Not only can the entity "be at" more than one place at one time, it is "at" all the spatial locations through which its episode passes during any of the time locations its episode encompasses.

To an outside observer that possesses an Ephemeral TI perspective, spatial positions within an episode are indeterminate. An episode ordinarily encompasses a span of time coordinates, but the measurement interaction prevalently takes place over just a fraction of that span. An observation is the concluding interaction of an episode and represents a "sampling" of some portion of that episode. Because an entity's footprint covers its entire present episode concurrently, the measurement interaction could potentially observe any part of the episode, though with differing degrees of likelihood. The initiating and concluding interactions, as boundary events, are definite constraints on an episode's possible path. The spacetime coordinates of the initiating and concluding interactions are, by definition, the only points within an episode that an observation can occur. Information from direct observation is not available at the other spacetime coordinates within an episode. Hence, an observer cannot ascribe a trajectory to an episode.

Arrangements of both spatial and temporal locations can be described with coordinate systems. While it is clear that the relative spatial coordinates of two different locations do not imply any order in which they must be inhabited, it is automatically assumed (from an Ephemeral TI perspective) that relative temporal coordinates inherently do confer an order of inhabitation. By direct contrast, for Episodic TI, there is no inherent ordering in which temporal locations within an episode are progressively inhabited. All of the events within an episode are concurrent, including the entity's inhabitation of separate spacetime coordinates. Separate events can thus be at different time coordinates within an episode, yet not be separated in the entity's experiences (analogously to how one's hands are arranged on the left and right, but their experiences are not ordered so that the sensations felt by one hand inherently precede the sensations felt by the other.)

An episode's path, from initiating to concluding interactions, is the worldline of the entity inhabiting that episode. Even an observer that directly observes the initiating and concluding interactions of an episode will not have traversed the same worldline as the entity between those interactions. The observer thus cannot directly determine the entity's elapsed proper time between interactions, much less its complete path. Prior to a concluding interaction, i.e. an observation, the entity's position is indeterminate, and the span of the entity's episode is unrestricted. The span of an episode which terminates in an observation can even encompass events with time coordinates that appear, from the observer's reference frame, to be later (i.e. in the future) than the time coordinates of the observation event. However, relativity clearly shows that only events on an individual world line can be ascribed an absolute time ordering. Temporal labels such as "earlier" and "later" attributed by an observer to perceived time coordinates of events that are outside that observer's world line are not meaningful outside of that observer's subjective experiences. Hence, Episodic TI provides a material mechanism to account for the QM formalism's prediction that there is a non-zero probability of finding an unobserved entity at any location.

Although Episodic TI might initially appear far-fetched, if not preposterous, a variety of considerations illustrate that Episodic TI is, in fact, improbably plausible. Consider first a photon propagating through a vacuum. From the photon's perspective, moving at the speed of light, time does not pass. The photon could traverse huge expanses of space and time between interactions, encompassing vast spans of time coordinates. Yet, the photon would "experience" all of that span without the passage of time in its reference frame, i.e. concurrently. The photon clearly experiences Episodic TI. This is a highly limited analogy, though. Entities that move at sub-light speeds will not experience this degree of time dilation. The relativistic effects of light-speed travel do not demonstrate the validity of Episodic TI alone, but they do provide an instructive means of investigating Episodic TI with a well-accepted phenomenon.

Consider next that virtually all physical entities occupy some expanse of space, since even "point" entities are generally thought to not be localizable to a point location in space, due to uncertainty constraints. Bear in mind also that traditional QM wave functions have non-zero probability distributions extending spatially to vast distances, and even, in theory, infinity. General relativity shows that the curvature of space varies with the distribution of matter. Variations in the curvature of space produce gravity and variations in the rate of time passage. Other than by an exceptional coincidence, two different locations in space will not experience absolutely identical gravitational fields for more than a moment (if then), even though they are very close together. Differences in the rate at which time flows at different locations across an entity's extent may seem to be almost trivially small, but any variation is difficult to reconcile with the precepts of Ephemeral TI. Furthermore, the variation may not be small after all. As time passes, different portions of an entity with a substantial spatial extent or in proximity to a great matter density could experience significantly different "present" moments. The QM orthodoxy is thus seen to already represent physical entities with extended wave functions that credibly encompass more than one point in time in what they describe as the same present moment (irrespective of the orthodox presumption that the entire entity experiences a single ephemeral present).

In a related manner, for an entity with a nonzero rest mass, the uncertainty in its momentum is commensurate to the entity's motion being comprised of a range of velocities. Time passes at different rates for reference frames moving with different velocities. Therefore, even if the entire entity initially had a present moment at a single location in time, its "present" will subsequently spread over a span of time coordinates. Hence, an application of basic special relativity to traditional QM models of physical existence lends plausibility to a non-ephemeral present moment, as did the immediately preceding applications of general relativistic considerations.

### EPR Paradox and a Redefinition of "Local"

Episodic TI furnishes a rational, local basis for the EPR paradox. The prototypical EPR pair of entangled particles share a superposition of quantum states, in episodes extending through time from their shared inception. The particles' ongoing present episodes conclude when one or the other experiences an observing interaction. An entanglement ending interaction must be a "non-local" action if the "present" moment is ephemeral. But if the "present" is episodic, then the concluding and initiating interactions of the measured particle are concurrent. The entangled particles' episodes share the same initiating interaction, so that the unmeasured particle's initiating interaction is also concurrent with the concluding interaction of the measured particle. Hence, the measurement interaction of either entangled particle is "local" to the other particle, irrespective of the spatial separation between them at the time location (in the measurement reference frame) that the measurement is recorded.

Any portion of a first entity's episode can interact "locally" with a second entity when any portion of the first entity's episode has a spatially local interaction with any portion of the second entity's episode. In this way, separate entities at great divides can directly and instantaneously interact locally. The key is that locality cannot be examined purely as a spatial quality. Even when the two entities are divided by a vast spatial gulf (at some time coordinate in either entity's frame of reference), their episodes will generally encompass a range of spacetime coordinates, including some which may be very close together. As a result, two entities could presently be both in immediate proximity and at a great distance.

## The Measurement Problem and the Quantum/Classical Boundary

The measurement problem is simply not a problem with Episodic TI, and deciphering how a wave function collapses upon observation is immaterial. Traditional QM interpretations hypothesize that a wave function collapses, globally and instantly, upon measurement. But relativity has no provision for constructing universal time foliations with which to define a global instant. The Many-Worlds Interpretation of QM (Everett, H., 1957; DeWitt, B. S., and Graham, N., 1973) is one notable result of attempts to resolve this quandary. The measurement problem is an inherent property of observer-actuated interpretations, and is actually an artifact of an ill-defined quantum/classical boundary.

The quantum/classical boundary issue exemplifies traditional QM interpretations' conceptual problems. If physical existence only exhibits Ephemeral TI, quantum and classical phenomena are inextricably distinct, and the attendant boundary problem is unavoidable. Solutions to the means of transition between the disparate quantum and classical regimes tended to be ad hoc reactions, rather than theories evolved from first principles. Whether the boundary is accounted for with statistical arguments, decoherence effects (spontaneous and/or environmental), or new interpretations of QM, the classical and quantum realms are still qualitatively incompatible. The aforementioned types of boundary explanations are used to give reasons for how a classical "appearance" arises from a quantum reality. But the transition remains uncertain, both in scope and character. Additionally, the previously safe theoretical zone, between the macroscopic and the microscopic, in which to sequester the transition is being rapidly intruded upon by the aforementioned experimental advances investigating mesoscopic quantum phenomena.

For Episodic TI there is no boundary problem. Consequently, it isn't necessary to justify the emergence of the classical realm by surmising a statistical averaging or conjuring a decohering environmental effect. Pending verification, this is among the foremost accomplishments of Episodic TI. The appearance of a boundary arises because sentient beings consist of such considerable concentrations of matter that they can only experience time ephemerally. What is perceived as a boundary is simply the inability of an Ephemeral observer to inhabit time Episodically. The Ephemeral TI nature of the classical world is due to the great density of entities in "normal" matter. Very frequent inter-entity interactions, and hence short episode spans, result. As discussed earlier, the application of calculus theory to physics is based on taking an interval of summation to the limit of zero duration (***dt***). The merit of calculations based on an interval of exceptionally short, but not zero, duration is unsettled. The author expects that, for the precision limits and uncertainty ranges of most measurements, minute intervals are likely to be valid approximations to the mathematical ideal embodied by Ephemeral TI. Granted, this approximation is a relatively major divergence from the theoretical assumptions which underpin familiar physics practices. Upon further reflection though, it is not as extraordinary as it may at first seem, given that approximations are at the heart of innumerable physics techniques, and are the essence of computer simulations and numerical calculations. How the degree of accuracy of this approximation varies with differing episode lengths, and how it compares to experimental results across a range of measurement forms are issues that need to be resolved. This will be an important area for future investigations to confirm or disprove, but it is beyond the scope of the present considerations.

A previously unattainable depth of comprehension is gained through Episodic TI's resolution of the boundary problem. No longer are there separate realms describing irreconcilable, and yet indivisible realities. Humans experience a classical realm because they consist of crowded conglomerations of practically innumerable entities, which are interacting nearly constantly. The Ephemeral TI appearance of the classical realm is an intrinsic product of the Episodic TI nature of entities, when highly concentrated.

## CAUSAL LOOP CREATION EXPERIMENT

### Background

Moving beyond conjecture and discussion, the Causal Loop Creation Experiment (CLCE), which suggests that the Ephemeral TI perspective may be untenable, will now be described. The CLCE is not only a logical argument that explores the consequences of known quantum phenomena, but also an experimental proposal that appears to be physically realizable. The logical line of reasoning alone places the Ephemeral perspective in doubt.  Moreover, physical execution of the CLCE, if successful, would both imply that Ephemeral TI is unsound, and indicate that interpretations incorporating Ephemeral TI, such as the Copenhagen and Many-Worlds Interpretations, are problematic.  By contrast, the plausibility of the Episodic TI perspective is bolstered by the logical argument, and would be greatly strengthened by a successful physical achievement of the CLCE.

The proposed experiment utilizes entanglement to explore a scheme for creating causal loops.  If Ephemeral TI is factual, causal loops will be created.  Virtually nothing is more antagonistic to the accepted conceptions of the nature of physical existence than is the prospect of a causal loop.  The logical argument of the gedanken CLCE shows that in order for one to cling to the presumptions of Ephemeral TI, one will also be compelled to accept causal loops.

Stimulus for the proposed experiment was provided by Percival's (Percival, I. C., 2000) Figure 3: "Spacetime diagram of the double Bell experiment, with photons in optical fibres."  Percival stated that a causal loop between separate reference frames of measurement angle choices and subsequent measurements would be created, unless one were to accept assumptions against causal loops (and backward causality) and thus be forced to exclude one of the non-local interactions.  Percival (Percival, I. C., 1998) concluded that dynamics were dependent on the Lorentz frame, which corresponds with Hardy's (Hardy, L., 2001) theorem that measurement dynamics, in flat spacetime with local realism and consistent with quantum theory, need a special Lorentz frame.

### Proposed Causal-loop Experimental Set-up

The experimental set-up schematically diagrammed in Figure 1 is (with minor, immaterial differences) as described in Percival's (Percival, I. C., 2000) Figure 3.  Figure 2 depicts a schematic of the proposed CLCE, which is closely related to Percival's double Bell experiment, albeit with critical modifications.  In Figure 2, all of the events in either reference frames A or B which are involved in the execution of the CLCE are spacelike separated from any of the events involved in the CLCE in the other reference frame.  (The relative displacements along the space and time axes are not depicted to scale in Figures 1 and 2, with the spatial separations being illustrated as relatively compacted.)  S and S′ are each sources of entangled photon pairs that propagate along optical fibers.

In Figure 2, two optical fibers lead from each photon source, one fiber from each source connecting to each reference frame.  The fibers are connected to the observers' reference frames so that in one reference frame an "earlier" fiber from one source intersects with that reference frame at an earlier time coordinate, as measured in that reference frame, than does a "later" fiber from the other source, while in the other reference frame the order of arrival of the fibers from the sources is reversed.  In frame A the receipt of S′ photons precede the receipt of S photons, while in frame B the order is reversed.   The following event sequence will then occur in each of reference frames A and B:

1. Decision on the type of the first measurement.

2. Interaction of "earlier" photon(s) with first measurement apparatus.

3. Decision on the type of the second measurement.

4. Interaction of "later" photon(s) with second measurement apparatus.

Specification of the relative time ordering of events in the two reference frames is pivotal to the CLCE's effectiveness.  It is not possible to ascribe an absolute (i.e. agreed upon by all possible observers) time ordering to events that are spacelike separated, at least without a preferred Lorenz frame such as discussed by Hardy (Hardy, L., 2001).  In fact, the time ordering is so uncertain that for spacelike separated observers moving relativistically in relation to each, when measuring a pair of entangled entities it is possible that "…both observers equivalently claim to trigger the collapse first!" (Zbinden, H., Brendel, J., Gisin, N.  and Titel, W., 2001) The creation of causal loops, for Ephemeral TI, will be seen to result if it can be prescribed that the measurements $A_2$ and $B_2′$ are executed before the measurements $B_2$ and $A_2′$, respectively, are executed.  An unambiguous realization of this will be accomplished by utilizing a delayed choice measurement protocol.

As discussed by Wheeler (Wheeler, J. A., 1978), the phenomenon of delayed choice measurements is implicit in orthodox QM.  Delayed choice measurements has since been investigated and confirmed (Zbinden, H., Brendel, J., Gisin, N. and Titel, W., 2001; Kim, Y.-H., Yu, R., Kulik, S. P., Shih, Y. and Scully, M. O., 2000) both for photons and massive particles. These delayed choice results have been confirmed both for spatial delays (where the choice is made when the spatial location of the measured entity is at a position that is distant from contact with the measurement apparatus) and for time delays (where the time that the choice is made is after when the entity was interacting with the measuring apparatus). Of course, relativity teaches us that these are not fundamentally different forms of delay, just different ways of describing the delay. Franson's (Franson, J. D., 1985) discussion of the delayed choice effect is particularly enlightening wherein he explains that "…the outcome of an event is not determined until some time after the event has already occurred." It has become clear that there is not, in principle, any absolute limit on the degree of this delay, other than is presented by technical constraints. The entire phenomenon of delayed choice measurements is also at the heart of the unresolved issues addresses by the present paper, and is more intuitively explained by Episodic TI (in the opinion of the author) than by other interpretations of QM. The CLCE will employ a delayed choice measurement protocol to accomplish a specific time ordering between events that would otherwise be spacelike separated and thus could not be assigned an unambiguous time ordering.

The events 2 and 4 described above are occurrences of the physical interaction of the relevant photon(s) with measurement apparatuses. As shown in the delayed choice research discussed, a measurement is not necessarily "executed" when the physical interaction with the measurement apparatus occurs, but may be delayed. Here, execution of events 2 through 4 in each reference frame are delayed sufficiently to enable a timelike connection from the event 4 interactions in each reference frame to reach the opposite reference frame prior to the execution of the event 2 measurement in the opposite reference frame. In this way, the second measurement in each reference frame is executed prior to the first measurement in the other reference frame. After the first measurement in a reference frame is executed, the result of the first measurement is communicated via a classical (timelike) link to the decision on the type of the second measurement in that reference frame. In each observer's reference frame, the prescribed measurement protocol thus produces the following sequence of measurement events:

1. Decision on own type of first measurement.

2. Execution of other observer's second measurement.

3. Execution of own first measurement.

4. Determination of type of own second measurement according to applicable decision rule applied to own first measurement result.

The delayed choice measurements and the timelike connection between reference frames does remove the CLCE from that class of experiments which close all possible locality loopholes, but this does not alter the substance or the significance of its results. Even though the measurement executions are delayed sufficiently to enable the timelike connection, all of the actual physical interactions of the photons with either reference frame are spacelike separated from all of their entangled counterparts' interactions with the other reference frame. Only the determinations of the measurements' results involve a timelike connection between the reference frames.

In the CLCE, a critical alteration to the experiment of Percival (Percival, I. C., 2000) is that the decision to be made involves a choice among different types of measurement, rather than a choice between settings of measurements of the same type. In Percival's double Bell experiment, as depicted in Figure 1, $A_1 / B_1{}'$ is the setting of the angle of spin or polarization measurement, while the +/– at $A_2 / B_2{}'$ represents the recording of the spin or polarization. In the CLCE, the choice in types of measurement is between an entanglement confirming measurement and an entanglement demolishing measurement.

The entanglement confirming measurement is to be devised and executed with care. Each source produces pairs of entangled photons that are entangled in shared superpositions of states. An entanglement confirming measurement is actually confirming that a photon is in a superposition of states, thereby indicating that it is also entangled. While a number of factors may improve the ease or certainty with which entanglement is confirmed, one specific property is clearly necessary: Conducting the measurement that confirms entanglement of one photon of an entangled pair should not necessarily terminate the other's superposed state, so that a similar entanglement confirming measurement can be conducted on it in the other reference frame. This will then be a form of a Quantum Nondemolition (QND) measurement. QND generally refers to a measurement which does not destroy the quantum entity being observed, so that it may later be re-observed. In this case, the QND character of the entanglement confirming measurement stems from the stipulation that it does not demolish the superposition of states of the photon that is the entangled counterpart to the photon being measured. Hence, the entanglement confirming measurement is also an entanglement preserving measurement (or, to be more precise, a measurement preserving of the superposition of states of the counterpart photon that is entangled with the measured photon). The measurements described as entanglement confirming are actually confirming that the photon exists in a superposed state, which is entangled with its counterpart photon's superposed state. The measurements described as entanglement demolishing are actually demolishing of a photon's superposition of states, and are hence demolishing of the superposed states of both the measured photon and its entangled counterpart. A result described as entanglement missing actually fails to find the presence of a superposition of states, which indicates the entanglement is absent (and implies that the entanglement has been demolished). Measurement protocols which are completely QND, both preserving and confirming of all photons and their states (whether entangled or not) would be ideal, but may not yet be practicable.

Decisions by observers conducting the measurements in the two reference frames determine if a causal loop is created. The decisions by observers A & B (of reference frames A & B, respectively) are between entanglement confirming (preserving) and entanglement demolishing measurements. The decision on the type of the measurements at $A_2 / B_2^{(')}$ are made at $A_1 / B_1^{(')}$, respectively. There are two rules controlling the decision of which type of measurement to make at $A_1$ and $B_1'$. The first rule is that the measurement action should be the same as the observation, while the second rule is that the action should be the opposite of the observation. For the first rule, when the first observation result is entanglement confirming, then the second type of measurement is entanglement preserving; whereas when the first observation result is entanglement missing then the second type of measurement is entanglement demolishing. In the case of the second rule, when the first observation result is entanglement confirming, then the second type of measurement is entanglement demolishing; whereas when the first observation is entanglement missing, then the second type of observation is entanglement preserving.

### CLCE Execution and Results

The purpose of conducting the Causal Loop Creation Experiment is to generate and identify causal loops that will arise from known quantum phenomena, if the presumptions of Ephemeral TI are correct. The causal loops are revealed by conflicts between earlier and later actions or observations arising in either, or both of, reference frames A or B. Appendix A to the present article provides a detailed explication of the CLCE's three scenarios, and their variations; while appendix B describes schemes for physically realizing executions of the entanglement confirming/preserving measurements described herein.

### Scenarios # 1 & # 2 – Causal Loop Non-Creation

There are three general scenarios to analyze. In a first scenario, both observers' measurement decisions are made according to the first rule – action same as observation. For purposes of brevity, in both the diagrams and the following discussions, YES indicates an entanglement preserving action or an entanglement confirming measurement, and NO indicates an entanglement demolishing action or an entanglement missing measurement. In other words, following a YES action or measurement, entanglement (and hence superposition of states) continues, and following a NO action or measurement, entanglement (and hence superposition of states) does not continue. Hence, action same as measurement means that when a first measurement has a YES result, then a YES action is taken and similarly for a NO result and action; and, accordingly, action opposite of measurement means that when a first measurement has a YES (NO) result then the second action taken is, respectively, a NO (YES) type of measurement. Inspection of Figure 3 reveals that the first scenario does not result in the creation of causal loops, when it is assumed that its implementation is ideal, i.e. no unintended decoherences occur.

In the second scenario, as shown in Figure 4, both observers make their measurement decisions according to the second rule – action opposite of observation. Action opposite of measurement means that when a measurement has a YES result, then a NO action is taken and similarly for a NO result and a YES action. Inspection of Figure 4 reveals that the second scenario also does not result in the creation of causal loops, when it is assumed that its implementation is also ideal, though which result $A_2'$ or $B_2$ is YES and which is NO may depend on the relative elapsed time periods required by the two observers to go through their successive decisions and measurements.

Scenario # 3 – Causal Loop Creation

In the third scenario, one observer (A in this case) makes measurement decisions according to the first rule – action same as observation; while the other observer (B) makes measurement decisions according to the second rule – action opposite of observation. In the prior analyses of scenarios # 1 & # 2, as well as the following analysis of scenario # 3, the sequence in which actions or measurements in one reference frame are evaluated may not coincide with those actions' or measurements' actual sequence of occurrence. The existence of a causal loop is found by determining the consistency of actions and measurements in their order of occurrence as experienced by an observer in either individual reference frame, regardless of the sequence of evaluations employed for the present analysis.

$1^{st}$ and $2^{nd}$ cases

There are three separate cases to consider for the third scenario. In a first case, observer A's first action ($A_1'$) is a NO action. In a second case, observer B's first action ($B_1$) is a NO action. Neither the first nor second case, when the implementation is ideal, results in the creation of causal loops, and hence will not be discussed in greater detail here (see appendix A for the complete discussion).

$3^{rd}$ case

In the $3^{rd}$ case of scenario # 3, the first action in both reference frames is YES. However, the first observation in either reference frame is not necessarily YES. An entanglement preserving measurement will not produce an entanglement confirming result if the entanglement has already been demolished when the measurement is made. In order to simplify the discussion, without loss of generality, the two observers' reference frames are held at static relative positions throughout the CLCE, and it follows that they will register the same elapsed proper time throughout. Furthermore, for the periods of time required for all of the photons to interact with the measurement apparatuses, all of the interaction events in either reference frame are spacelike separated from all of the interaction events in the other reference frame. The delayed choice measurement protocol ensures that one reference frame's second measurement of a photon's entanglement is executed before the counterpart photon's entanglement undergoes the other reference frame's first measurement. The preservation of a photon's entanglement at either reference frame's first measurement will therefore depend on the type of the measurement made at the other reference frame's second measurement. As described earlier, the sequence in which events are considered need not correspond to the sequence that events actually occur in either reference frame. Two approaches are employed for the sequences of considerations. In a first approach variant, the result of the second measurement in either reference frame A or B is postulated, and the following events are then determined; whereas in a second, alternate approach variant, the result of the first measurement in either reference frame A or B is postulated and the following events are then determined.

As explicitly demonstrated in Appendix A, the $3^{rd}$ case event sequences show that when both observers initially act to conduct an entanglement confirming measurement, subsequent events will conflict. In the first variants of the $3^{rd}$ case, the assumed result of a measurement ($A_2$ or $B_2'$) is incompatible with the prior action ($A_1$ or $B_1'$, respectively) that determines the type of the assumed measurement. When the assumed measurement result is switched, and the event sequence is then determined anew, one of the alternate variants of the $3^{rd}$ case is the outcome. Inspection of the alternate variants of $3^{rd}$ case reveals that they all manifest contradictory events also. Therefore, in a reality explained by Ephemeral TI, all of the variants of the $3^{rd}$ case of scenario # 3 of the CLCE will unequivocally create causal loops.

### Significance of the Causal Loop Creation Experiment

Among the repercussions of the CLCE are its highlighting of the inadequacies of the Copenhagen, Many-Worlds, and other O-AI's. These interpretations were conjectured to account for seemingly inexplicable quantum phenomena. The causal loops definitely created in scenario # 3 (and likely to be regularly occurring in scenarios # 1 and # 2) are a consequence of the various O-AI's inherent incorporation of the Ephemeral TI perspective. The root of the defect is Ephemeral TI's erroneous description of physical existence. Though the particulars of the various O-AI's facades are not the immediate source of their inadequacies, resolving these inadequacies with Episodic TI eliminates the motivation for creating these interpretations. Whether grounds remain to justify their continuance is now the primary question for the expounders of O-AI's.

Further examination of the scenario # 3 alternate variants reveals that a direct consequence of the Ephemeral TI presumption is the creation of causal loops. Converting the test assumptions (step **3** of the event considerations of the 3rd case of scenario # 3) of the alternate variants to coincide with later events is impossible with Ephemeral TI, since a past moment in time is inalterably sundered from later points in time. Episodic TI does not prevent these alterations, though. With Episodic TI, a pair of entangled photons' present episodes share their point of initiation so that, as long as the entanglement remains, any event experienced by one photon is concurrent with any event experienced by the other. The effect for the present experiment is to remove the Ephemeral TI barrier to revisions of "past" events, because in Episodic TI the events within an episode are not actually "past".

So as to ensure that the meaning of a non-ephemeral "present" episode is clear, it is important to note that Episodic TI is not related to the concept of a closed timelike (or null) curve that arises in certain discussions of General Relativity. Among those discussing issues related to closed curves are Novikov, Frolov, Thorne, Deutsch, Hawking and numerous others. Visser (Visser, M., 2003) presents a general survey and discussion of the various lines of investigation of closed curves. While these curves, if they exist, could give rise to causal loops, this is a result of the world lines they describe intersecting with themselves, and hence being "closed". However it is conjectured that these curves may (or may not) be created, and whatever their potential effects or quandaries, the presumptions of Ephemeral TI are integral characteristics of these conjectures. Regardless of whether or not a null curve is deemed traversable (i.e. a prior point in time can be revisited), these conjectures assume that each point along such a curve is considered to be a separate "present moment" from the immediately prior and following moments, in direct contradiction to the underlying concept of Episodic TI. While it does not necessarily preclude or alter theses discussions of closed timelike curves, Episodic TI might impact on their eventual resolutions, though that is a discussion beyond the scope of the present paper.

Episodic TI's capacity to modify the "past" both clarifies the analyses of scenario # 3 and provides an impetus for adopting the Episodic TI perspective. In applying Episodic TI to the CLCE, the "past" can be modified because, within an episode, an event at an earlier time coordinate is concurrent with an event at a later time coordinate. Accordingly, though $B_2$ occurs at an earlier time coordinate than $B_2'$ in reference frame B, $B_2'$ is also concurrent with $A_2'$ which occurs earlier than $A_2$ in reference frame A, plus $A_2$ is concurrent with $B_2$. Clearly, the events in the CLCE cannot be ascribed an absolute order of occurrence. With Episodic TI, these events unfold in a cyclic progression that nullifies the possibility of forming a causal loop. For Ephemeral TI, the past is inviolate and causal loops cannot be avoided. Stepping through the variants of scenario # 3 event-by-event will explicate the resolution provided by Episodic TI.

<u>Verification of Test Assumptions</u>

The following evaluations demonstrate how the outcomes of the $3^{rd}$ case variants can be definitively resolved with Episodic TI, even though the events are seemingly interdependent in an unending loop of consequences. The solutions are found by taking into account all of the possibilities a measurement can produce, and then following the sequence of events that ensue after a given assumption. When conflicts between succeeding events and an initial assumption happen, there are two options: either the assumption changes to conform with later events, or the assumption is shown to be erroneous when it cannot be reconciled with the later events it would have produced. Fortunately, there are only four measurements made in the CLCE, with two possible results for each measurement, so that just eight total variants can arise. In the first variants, the test assumption is made about the result of the later measurement in each reference frame. In the alternate variants, the test assumption is made about the result of the earlier measurement in each reference frame. When the test assumption is that the result of a measurement =YES, but later events show the action controlling the type of the test assumption measurement is NO, then the result of the test assumption measurement can be changed to =NO in a subsequent cycle, since it is possible to demolish an existing entanglement. When the test assumption is that the result of a measurement =NO, but later events show the action controlling the type of the test assumption measurement is YES, then the result of the test assumption measurement cannot be changed to =YES, since it is not possible to ***un***demolish an entanglement. In this instance, the test assumption is seen to be erroneous and must be changed to =YES (if it is assumed that any unintended entanglement demolishment is not the cause), and the sequence of events are then to be reevaluated from the beginning.

For the first variants, the evaluation sequences plainly show that all of the test assumptions are in conflict with earlier events in the reference frame of the observer making the test assumption. For the first variants A or B =YES (test assumptions $A_2$ or $B_2'$ =YES, respectively), the eventual effect that $A_1$ or $B_1'$ =NO, respectively, will change $A_2$ or $B_2'$ to =NO, respectively, after one cycling of events. the next cycle of events will then change the results of all the measurements to NO, and no further alterations to the measurement results, and hence to the actions, will occur after the completion of that cycle. Potential causal loops are thus eradicated by the demolishing of all entanglements. For the first variants A or B =NO (test assumptions $A_2$ or $B_2'$ =NO, respectively), the evaluation outcome that $A_1$ or $B_1'$ =YES, respectively, proves that the test assumption is erroneous, since the assumed test results =NO would not have arisen from their precipitating YES types of measurements. (The =NO result of $B_2$ or $A_2'$ is a consequence of the test assumptions in these variants. Neither $B_2$ nor $A_2'$ will demolish their respective entanglements, since their types of measurements are both =YES.) Although a YES result can change to a NO result, the reverse is not possible. Correcting the test assumption (by altering $A_2$ or $B_2'$ to =YES, respectively) and reevaluating the sequence of events from the beginning just repeats the first variants A or B =YES evaluated immediately above. Consequently, the outcomes of all of the first variants that are possible will be findings that all entanglements are missing, for both the earlier and the later measurements in both reference frames.

Applying the test assumption appraisal principle outlined above enables a straightforward evaluation of the alternate variants also. For the alternate variant A =YES, the test assumption $A_2'$ =YES changes to =NO after an initial cycle of evaluations, and the second cycle of event evaluations are then $A_1$

=NO, $A_2$ =NO, $B_2$ =NO, $B_1{}'$ =YES, and $B_2{}'$ =NO. Despite $B_1{}'$=YES in the second cycle, $B_2{}'$=NO since the S′ photons' entanglements had already been demolished in the initial cycle. The end result is that the entanglements of the S and S′ photon pairs have both been demolished (all measurements =NO).

For the alternate variant B =YES, the test assumption $B_2$ =YES changes to =NO after an initial cycle of evaluations. The ensuing second cycle of event evaluations are then $B_1{}'$ =YES, $B_2{}'$ =NO, (once again, despite $B_1{}'$ =YES in the second cycle, $B_2{}'$ =NO since the S′ photons' entanglements had already been demolished in the initial cycle), $A_2{}'$ =NO, $A_1$ =NO, and $A_2$ =NO. The end result is that the entanglements of the S and S′ photon pairs have once more both been demolished (all measurements =NO). For Episodic TI, the result of a measurement is capable of being altered without a "future" event causing a change in a "past" event because the measurements of the entangled photons are taking place within a single intact episode and hence are concurrent. The measurement result can change without retroactively altering the past because the measurement result ***is not in*** the past. By contrast, for Ephemeral TI the two measurement events have distinctly separate positions in time that have an absolute future/past relationship, and a causal loop is inevitable.

Turning to the alternate variant A =NO, it is seen that the test assumption $A_2{}'$ =NO would not come about from a combination of the postulated measurement action $A_1{}'$ =YES and the ensuing measurement result $B_2{}'$ =YES, since an entanglement confirming measurement of an intact entanglement would have returned a result of =YES. The test assumption is therefore erroneous, and must be switched. Correcting the test assumption (by altering $A_2{}'$ to =YES) and reevaluating the sequence of events from the beginning just repeats the alternate variant A =YES evaluated previously. Similarly, for the alternate variant B =NO the test assumption $B_2$ =NO would not come about from a combination of the postulated measurement action $B_1$ =YES and the ensuing measurement result $A_2$ =YES. This test assumption is therefore erroneous also, and it must be switched as well. Correcting this test assumption (by altering $B_2$ to =YES) and similarly reevaluating this sequence of events from the beginning just repeats the alternate variant B =YES evaluated previously.

Establishment of the complete set of CLCE results by evaluating all possible test assumptions for the measurement events from an Episodic TI perspective has revealed that the final outcome of the CLCE, when correctly executed, will be a conflict resolution that results in all measurements finding that the entanglement is missing. This self-directed disappearance of all entanglements is termed here "autonomous decoherence".

Definitive verification that the CLCE's outcome is a departure from what can be explained by Ephemeral TI can be confirmed by initially checking disconnected portions of the CLCE apparatus. Each of the four measurements should register the expected confirming (or missing) results when they are conducted individually, and in combinations when one or the other (but not both) of the sources are operating. The mere fact that both sources are emitting entangled photons in tandem should not change these results, at least according to the precepts of Ephemeral TI. The interleaving of the present episodes in the CLCE will alter these results, however, and thereby provide a persuasive affirmation of the Episodic TI hypothesis.

The CLCE is well suited for enlightening comprehension of O-AI's, because the consequences of observers' actions are intimately involved in the execution of the CLCE. As described in appendix B, recent experimental developments make it clear that the CLCE is not only a reasonable logical gedanken experiment, but that it is also physically feasible to conduct at the present time. If reality is dependent on the actions of observers to "come into existence", the CLCE will be an important test of the soundness of O-AI's, unless Hardy's preferred Lorentz frame can be conjured forth. Even so, should a preferred Lorentz frame be eventually divined, the consequences of the CLCE for Ephemeral TI will still stand.

Some might argue that the experimental set-up and preset coordination between observers of the CLCE is highly improbable, or at least of narrow applicability to the course of normal events. An appreciation of the breadth of applicability of the autonomous decoherence that occurs in the $3^{rd}$ case of scenario # 3 will assist in gauging this allegation. The question can be most simply stated by inquiring whether autonomous decoherence will also occur in scenario # 1 or # 2, or in the $1^{st}$ or $2^{nd}$ cases of scenario # 3. The only significant distinction between the $3^{rd}$ case of scenario # 3 and the others are the decisions on the types of entanglement measurements, and/or the choice of rules that control these decisions. These decisions are optional choices, which are not independently verifiable unless implemented by an automatic apparatus. Even then, particularly for decisions intended to execute entanglement preserving measurements, the actual measurement result can be altered, for example, by experimental error or a spontaneous decoherence.

For all three scenarios, when either of decisions $A_1'$ or $B_1$ are =NO, then interleaved present episodes are not present, and causal loops cannot result. However, when both decisions $A_1'$ and $B_1$ are YES in scenario # 1, then both $A_1$ and $A_2$ will be YES, also. It is possible for the photon traveling to $B_2$ to experience decoherence from environmental effects, for example, even though $A_2$ is YES. When the entanglement is lost in this way, then $B_2$ will be NO, of course, and hence $B_1'$ and $B_2'$ will be NO also. But when $B_2'$ is NO, then $A_2'$ must also be NO, and we then have a causal loop for scenario # 1 also. In a similar manner, it is evident that scenario # 2 can also produce a causal loop due to a "fortuitously" timed unintentional decoherence. Numerous experiments employing entanglement have documented that unintended decoherences occurs so readily that great care has to be exercised in their design and execution just to avoid it. Irrespective of the degree of care and control exercised, this phenomenon cannot be minimized for the CLCE. Beyond the question of whether or not an "infrequent" causal loop is a serious problem (chances are it is), there is also the issue of the ever-cycling event sequences due to the CLCE's interleaved present episodes. No matter how low the rate of decoherence can be reduced to, given that there are an unlimited number of opportunities for it to occur, the chance that it will, in fact, occur at least once is effectively unity. Thus, the real-world outcome of the CLCE is the production of causal loops in all three scenarios provided both decisions $A_1'$ and $B_1$ are YES, and provided the entangled photon pairs avoid unintended decoherence prior to at least a single complete circuit of the CLCE event sequences. Therefore, given this set of circumstances, execution of the CLCE is predicted to produce for all three scenarios the same end result as for the 3rd case of scenario # 3: both $A_2'$ and $B_2$ will return =NO measurements even though the photons leave the sources entangled.

## DISCUSSION AND CONCLUSIONS

Episodic TI's predictions for the CLCE result from the cyclic effect on events of the entangled photons' interleaved present episodes.  The cycling is present, but does not impact on events, unless the result of a measurement is altered from one cycle to the next (since the two first actions are predetermined, and the two second actions are conditional upon the results of the two first measurements).  As long as either entanglement is intact, it is possible to alter the result of any measurement which confirmed that that entanglement is intact.  Once both entanglements have been demolished, the results of the measurement events are set.  For the conventional conception of reality (Ephemeral TI), the logical gedanken CLCE shows this enables the creation of causal loops.  But with Episodic TI, this cycling of events is not a causal loop.  The events all occur concurrently, i.e. within the same present episode, and are hence contemporaneous (as well as indistinguishable and therefore inseparable to an outside observer).  Contemporaneous events don't create causal loops because there is no temporal ordering of events within an episode, and hence a "later" cause cannot produce an "earlier" result because ascribing terms such as "later" and "earlier" to events that occur at different time coordinates within an episode is not meaningful.

Unlike the vast majority of QM interpretations, Episodic TI presents a specific prediction that is readily testable by a physically realizable CLCE.  The results of the CLCE will be a strong test of the possibility of Episodic TI's validity, and offers the prospect of a definitive determination of the invalidity of Ephemeral TI based interpretations (including most all O-AI's).  It is possible to dictate that the decisions and or rules will always be made according to only those scenarios in which no causal loops would result.  In this way, the facade of an Ephemeral TI reality "safe" from causal loops may perhaps be maintained.  But even the most vociferous proponents of O-AI's would not attempt to preclude otherwise reasonable actions in order to safeguard the integrity of an interpretation, a course of action that is unacceptable to any self-respecting physical theory; and is a principal defect in the reasoning of Percival (Percival, I. C., 1999) wherein he declares that situations such as the  3rd case of scenario # 3 are forbidden by fiat since causal loops are, of course, verboten. Precluding the formation of causal loops in the CLCE is almost certain to be compulsory for any theory to retain legitimacy.

How widespread autonomous decoherence is – beyond the specific circumstances of the CLCE – is an open question. It could be that unless the measurement events are arranged as prescribed for the $3^{rd}$ case of scenario # 3 (or the causal loop creating unintentional decoherence events discussed for scenarios # 1 and # 2) that autonomous decoherence will not occur. In order to bring about autonomous decoherence, is it sufficient for separate reference frames to be connected by interleaved present episodes, or is it also necessary for the measurement events and decision rules to be carefully arranged (as in the CLCE)? The way in which causal loops can result from scenarios # 1 and # 2 is instructive for resolving this question. As long as interleaved present episodes are intact, it is possible for a "later" event to cause a change in an "earlier" event, if "later" and "earlier" are defined according to Ephemeral TI. There is nothing specifically unique about the form of measurements utilized in the CLCE. The strictures on the forms of measurements employed are obligated by the need to produce a controllable, verifiable, and repeatable experiment. Any of a wide range of physical processes that would not serve these three aims are still capable of functioning in the manner called for. The possibility of causal loops also emanating from scenarios # 1 and # 2 demonstrates that autonomous decoherence is probably not a limited artifact of a peculiar experiment.

It is the author's opinion that this process is more likely to be an inherent, pervasive aspect of physical existence, rather than a limited artifact of the particular arrangement of the CLCE. A more probable expectation (than a limited artifact) is that practically any situation that involves interleaved present episodes will induce autonomous decoherence, which would thereby imply its prevalence. If this is so, it could be indicated by the CLCE as well, in the instance where only one of $A_1'$ or $B_1$ are =NO. In either of these instances, though the interleaved present episodes are interrupted by the first NO measurement, they must still have been present at least momentarily until that first =NO measurement is executed. If $A_2'$ and $B_2$ both return =NO results even in this situation, it would be a strong indication that autonomous decoherence is more likely to be a pervasive phenomenon, since it will have then occurred even though the experiment's execution was specifically arranged to circumvent the present method of forming causal loops. Conducting the CLCE without the presence of the second measurement in either reference frame, and still finding that both observers' first measurement results were entanglement missing, would be an even stronger indicator that autonomous decoherence is pervasive.

To summarize, upon completion of the causal loop creation experiment set-up and transit of two pairs of entangled photons to their respective measurements, the Episodic TI perspective forecasts that the photons will have decohered autonomously, the entanglements will have ceased to exist, and the measurement results will all be "entanglement missing".

The amount of significance to attribute to autonomous decoherence is an open question. One deduction drawn here is a general assertion relating to the aptness of O-AI's. The cornerstone of O-AI's is that physical existence is definitively realized only through the action of an observer (regardless of however the nature of that observer is defined). The CLCE demonstrates that the mere modification of a decision rule by a single observer in either the first or second scenarios proves catastrophic for conventional (Ephemeral TI) interpretations of reality. For example, in scenario # 1 Ephemeral TI predicts that when both observers' first actions are =YES, then all the measurements are also =YES and, if so, all is well for Ephemeral TI and O-AI's. But changing just one of the decision rules, an action analogous to the paradigm at the heart of O-AI's, proves lethal to Ephemeral Time Inhabitation. Without Ephemeral TI, O-AI's are unnecessary. The root causes of the problem are the inadequacies of Ephemeral TI. That the problem can be solved by Episodic TI, and Episodic TI alone (given an assumption of strict special relativity) should, at minimum, give pause to those who expound O-AI's. Not only does Episodic TI provide intuitive (albeit fundamentally radical) explications of QM phenomena that O-AI's are incapable of; but the very leitmotif of O-AI's is seen to be the trigger for the creation of the causal loops which assail Ephemeral TI in the CLCE. Granted, this is essentially a judgment on the efficacy of employing an O-AI of physical existence. But without a problem to solve, and sizable dilemmas to overcome, the question is: are there any reasons (besides momentum) to continue jamming a round reality into a square O-AI paradigm?

In judging a hierarchy of repercussions of the CLCE, the author ascribes a lower import to its connotations for O-AI's, and a higher import to its refuting of Ephemeral TI. Although the CLCE demonstrates Ephemeral TI's deficiency, it does not conclusively prove that the only possible explanation is Episodic TI as envisioned herein. Whether or not physical existence is contained within a single time dimension, that existence can be regarded as occupying time in either an Ephemeral TI or a non-Ephemeral TI manner. Intriguingly, possible variations in the composition of the time dimension(s) do not preclude any of the elements of the Episodic TI hypothesis. The range of varieties of non-Ephemeral TI's are not limited in principle, outside of the need for congruence with the CLCE and the known results of QM measurements. The present Episodic TI hypothesis appears to be the least convoluted of the conceivable alternatives and is perhaps preferable for this reason, at minimum, in the absence of new evidence to the contrary. Initially, application of an Episodic TI perspective to investigations of other phenomena will offer avenues for corroborating Episodic TI's degree of validity. Eventually, and conceivably most importantly, Episodic TI should enable resolution of a great host of uncertainties in the understanding of quantum mechanics and related phenomena.

It is readily apparent that if Episodic TI proves correct, the ramifications of this hypothesis will be at least as extensive as its premise is radical. In fact, if it is found to be even partially valid, the impact will be widespread throughout physics, and even philosophy in general. Beyond validation by experiment, many challenges lie ahead for Episodic TI, not the least of which is the formulation of a rigorous theory of the physics of Episode evolution, which might be termed Episode Mechanics (or maybe Episodics). What seems clear is that, if Episodic TI is sound, an entity's episode is among the most elemental characteristics of physical existence. It even may well be that an episode should be considered the basic unit of tangible existence, and that studying an entity without cognizance of its episodic dimensions is an oversight comparable to ignoring its spatial dimensions. Inhabiting an episode is what it means to exist physically, the two are one and the same. The fruits of the application of the Episodic TI perspective to varying phenomena should yield both greater understanding of the present hypothesis, as well as new insights about the phenomena themselves. From issues as basic as the nature of wave/particle duality, to questions as cosmic as the contribution of virtual particles to the universal expansion rate, the eventual impact of Episodic TI is challenging to envision, but promises to be substantial.  In the author's opinion, the evidence for a serious consideration of the Episodic TI hypothesis is compelling, and a fair and exhaustive examination will eventually reward all.

## ACKNOWLEDGMENTS

The author would like to thank M. Visser for his generous discussions and advice, and A. Carlsson for his patient encouragement.

<u>*APPENDIX A*</u>

# ANALYSIS OF PROPOSED CAUSAL LOOP CREATION EXPERIMENT

## <u>Introduction</u>

The experimental set-up of Figure 3 from Percival (Percival, I. C., 2000) is schematically diagrammed in the present Figure 1. Figure 2 depicts a schematic of the proposed CLCE, which is fairly closely related to Percival's double Bell experiment, though with certain critical modifications. In Figure 2, all of the events in either of reference frames A or B that are involved in the execution of the CLCE are spacelike separated from any of the events involved in the execution of the CLCE in the other reference frame. (The relative displacements along the space and time axes are not necessarily depicted to scale in Figures 1 and 2, with the spatial separations being illustrated as relatively compacted.) S and S′ are sources of entangled photon pairs that propagate along optical fibers. A first critical modification to Percival's (Percival, I. C., 2000) double Bell experiment is that in the CLCE the decision to be made involves a choice among different types of measurement, rather than a choice between settings of measurements of the same type. A second critical modification is the employment of a delayed choice measurement protocol in each observer's reference frame that includes a transmission of the delayed choice registration of the other observer's second measurement outcome. These combined delayed choice measurements are executed so that the result of the second measurement in observer A's reference frame is registered by observer B in observer B's reference frame before observer B executes and registers observer B's first measurement, and vice versa. The sequence of observation events that involve each observer's reference frame are hence:

1. Decision on own type of first measurement.

2. Determination of other observer's second measurement result.

3. Determination of own first measurement result.

4.  Determination of type of own second measurement according to applicable decision rule applied to result of own first measurement.

5.  Determination (by other observer) of own second measurement result.

**Notation, Sequences, and Connections of Experimental Events**

In frame A the first decision is $A_1'$, the first photon/measurement apparatus interaction is $A_2'$, the second decision is $A_1$, and the second photon/measurement apparatus interaction is $A_2$; in frame B the first decision is $B_1$, the first photon/measurement apparatus interaction is $B_2$, the second decision is $B_1'$, and the second photon/measurement apparatus interaction is $B_2'$.  Entangled photons from source $S'$ are received in frame A after $A_1'$ and before $A_2'$, while entangled photons from source S are received in frame A after $A_1$ and before $A_2$.  Entangled photons from source $S'$ are received in frame B after $B_1'$ and before $B_2'$, while entangled photons from source $S'$ are received in frame B after $B_1'$ and before $B_2'$.  Finally, the result of the measurement at $A_2'$ is communicated over a classical link as input for the decision at $A_1$; and the result of the measurement at $B_2$ is communicated over a classical link as input for the decision at $B_1'$.

The actual execution of the measurements and decisions by the two observers are made according to the delayed choice protocols described in the present paper.  (Delaying the execution of each observer's first decision is unnecessary, but is not a hindrance either.)  The extent of the delay is sufficient for a timelike communication channel to be traversed between the reference frames so that observer A can conduct the actual execution of the $B_2'$ measurement after $A_1'$ and before the actual execution of $A_2$, and vice versa.

**Experimental Implementation**

SCENARIO # 1 – CAUSAL LOOP NON-CREATION

There are three general scenarios to analyze.  In the first scenario, both observers make their measurement decisions according to the first rule – action same as observation.  As shown in Figure 3, when both observers make the same first decision to demolish or preserve entanglement, the sequence of subsequent events, including the entangled interconnections between reference frames A & B, will be congruent with preceding events. Once the first YES/NO division has been made, succeeding events are in conformity with prior events.

When the observers make opposite first decisions, the situation is more complicated. The observer that initially acts to demolish entanglement enters the NO domain once the first measurement is made. The observer which makes a NO first action will return a NO first measurement result regardless of whether or not that photon was still entangled. In accordance with scenario # 1's decision rule, this observer's second measurement is also entanglement demolishing. That second measurement's demolishing of entanglement means that the once-entangled photon which is to be measured by the other observer's first measurement is no longer in a superposition of states. Since the measurement of the photon propagating along the longer fiber is executed before the measurement of its entangled counterpart photon propagating along the shorter fiber is executed, when the shorter fiber's photon is measured it has already had its superposition of states demolished. Hence, it is of no consequence if that observer has decided to conduct an entanglement preserving measurement, since it cannot confirm an entanglement which no longer exists. Therefore, the other observer then enters the NO domain also, and the subsequent events observed by either observer will thenceforth be compatible with prior observed events.

SCENARIO # 2 – CAUSAL LOOP NON-CREATION

In the second scenario, as shown in Figure 4, both observers make their measurement decisions according to the second rule – action opposite of observation. Once again, the execution of the measurement of the photons at $A_2$ and $B_2'$ prior to the execution of the measurement of their entangled counterpart photons at $B_2$ and $A_2'$, respectively, are the critical factors in determining whether or not the first measurement in either reference frame is measuring a photon that still possesses an entanglement to be confirmed (or demolished). Hence, to analyze the sequence of events in a given reference frame, the result of the second measurement in the other reference frame establishes whether or not the first measurement in the given reference frame is measuring a still entangled photon.

When the result of the second measurement in the given reference frame is NO (entanglement missing), then regardless of the first action in the other reference frame, the other frame's first measurement will be NO (entanglement missing). The second action in the other frame is then YES, and the second measurement in the other frame preserves the entanglement for the first measurement in the given reference frame. When the first action in the given frame is YES, in accordance with the scenario # 2 decision rule, the second action in the given frame is then NO, so that the second measurement in the given frame is entanglement demolishing and returns a result of NO, which is consistent with the initial assumption. Therefore, the events of the two reference frames are harmonious when the observers choose opposite first actions.

When the observers choose the same first action, the sequence of events will differ when they both choose YES than when they both choose NO. Turning initially to both first choices being YES, assume the result of the given frame's second measurement is YES also. The other frame's first measurement result will then be YES, and hence the other frame's second action will be NO, and so the result of its second measurement is NO also. The result of the first measurement in the given frame must also be NO, regardless of the first choice of a YES action. The second choice of action in the given frame is then YES, and the resulting YES measurement verifies the initial assumption.

When both first choices are YES, and the result of the given frame's second measurement is NO, the result of the other frame's first measurement must also be NO, regardless of the other frame's first choice being YES. In accordance with scenario # 2's decision rule, the other frame's second action is then YES and thus the other frame's second measurement result is YES. Since the given frame's first action is postulated to be YES, the result of the given frame's first measurement is likewise YES. The given frame's second action is accordingly NO, and thus the result of the given frame's second measurement is NO, which agrees with the initial assumption.

Which of reference frames A and B is the given frame and which is the other will depend on the reference frame of the individual making this determination. As discussed in the main body of this paper, it is not possible to ascribe an absolute ordering to events that are spacelike separated, at least without the discovery of a preferred Lorenz frame such as discussed by Hardy (Hardy, L., 2001). The crossing of the arrival times of the optical fibers between the two reference frames presents an ambiguous sequence of events.  It is conceivable that both second measurements could return NO results, and hence both first measurements would return NO results.  In this case, both second actions would be YES, though the execution of an entanglement confirming measurement is irrelevant, since the superposed states would have to already have been demolished for both second measurements to return NO.  Nevertheless, whichever of the observers' second measurements is determined to take precedence (according to a particular observer), once that determination has been made, the following events proceed without contradiction, at least to that observer.

Turning next to both observers choosing a first action of NO, the result of both first measurements will be NO.  Accordingly, both second actions will be YES and each of the second measurements in both frames will be YES also because they are executed prior to the first measurements' demolishing of their delayed counterparts' superposed states and hence, when executed, the second measurements are conducted on photons with intact superposed states.  Plainly, no conflict between earlier and later events arises in this case.  Having explored the various cases which events can follow in the second scenario, it is apparent that earlier results and actions do not conflict with those that follow for either observer, and hence no causal loops arise.

SCENARIO # 3 – CAUSAL LOOP CREATION

In the third scenario, one observer (A) makes measurement decisions according to the first rule – action same as observation; while the other observer (B) makes measurement decisions according to the second rule – action opposite of observation.  In the prior analyses of scenarios # 1 & # 2, as well as the following analysis of scenario # 3, the sequence in which actions or measurements are evaluated in one reference frame may not coincide with those actions or measurements actual sequence of occurrence in that reference frame.  The existence of a causal loop is found by determining the consistency of actions and measurements as they actually occur according to the observer in each individual reference frame.

1st Case

In the first case, observer A's first action ($A_1'$) is a NO action.  A's first measurement ($A_2'$), second action ($A_1$), and second measurement ($A_2$) are thus all NO also.  Since the $A_2$ measurement precedes the occurrence of the $B_2$ measurement, the result of $B_2$ is NO regardless of which action $B_1$ is. Since observer B follows the opposite decision rule, the action $B_1'$ and measurement $B_2'$ are YES. The preservation of the entanglement by $B_2'$ does not affect the NO result of the measurement $A_2'$, because an entanglement demolishing measurement has the same result whether or not the entanglement measured was still intact. Therefore, the first case does not directly give rise to a causal loop:

**SCENARIO #3 – 1st Case**

(Numerals Indicate the Sequence of Event Evaluations)

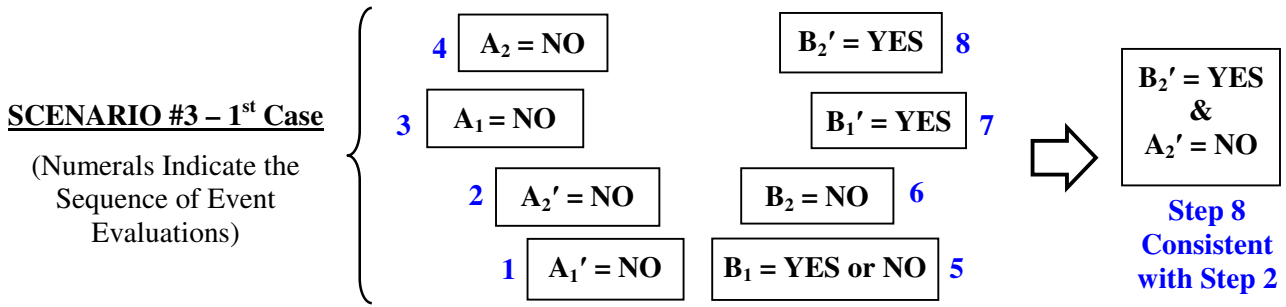

## 2nd Case

In the second case, observer B's first action ($B_1$) is a NO action, so that B's first measurement ($B_2$) is also NO. B's second action ($B_1'$) is then YES, and B's second measurement ($B_2'$) is thus YES also. $A_1'$ is taken to be YES, since the first case already treated $A_1'$ being NO. Because $B_2'$ is an entanglement preserving measurement, $A_2'$ returns a YES result, and hence $A_1$ is a YES action. $A_2$ has a YES result then, and reminiscent of the first case, the preservation of the entanglement for the measurement $B_2$ is irrelevant to the result of $B_2$. The second case measurement results and actions are consistent, and therefore the second case does not directly give rise to a causal loop:

**SCENARIO #3 – 2nd Case**

(Numerals Indicate the Sequence of Event Evaluations)

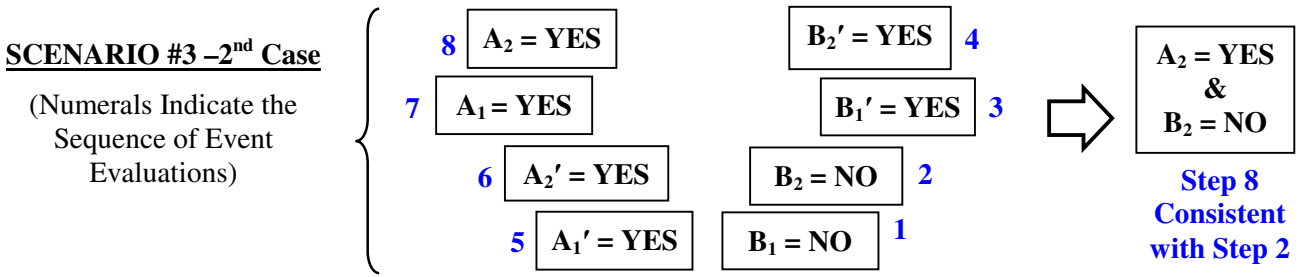

<u>3<sup>rd</sup> Case</u>

In the third case, observer A's first action ($A_1'$) is a YES action. The result of measurement $A_2'$ will depend on whether $B_2'$ is entanglement preserving or entanglement demolishing. The result of $B_2'$ will depend on whether $B_1'$ is YES or NO, which coincide with $B_2$ being NO or YES, respectively. When $B_1$ is NO, then $B_2$ is NO, and this is the second case already evaluated. $B_1$ is then YES, and the result of $B_2$ will depend on whether $A_2$ is entanglement preserving or entanglement demolishing. The result of $A_2$ will depend on whether $A_1$ is YES or NO, which coincide with $A_2'$ being YES or NO, respectively. There are a pair of variant options for evaluating the third case, *A* or *B*, according to whether the value of $A_2$ or $B_2'$ is assumed initially.

For variant *A*, assume first that $A_2$ is YES, so that $B_2$ is also YES. $B_1'$ is then NO, so that $B_2'$ is also NO. $A_2'$ must then also be NO, and thus $A_1$ is NO. But $A_2$ cannot be YES, when the action $A_1$ is to conduct an entanglement demolishing measurement at $A_2$. Assume instead then that $A_2$ is NO, so that $B_2$ is NO. $B_1'$ is then YES, so that $B_2'$ is also YES. $A_2'$ must then also be YES, and thus $A_1$ is YES. Again, the assumed result of $A_2$ conflicts with the action $A_1$ that determines the type of measurement conducted at $A_2$:

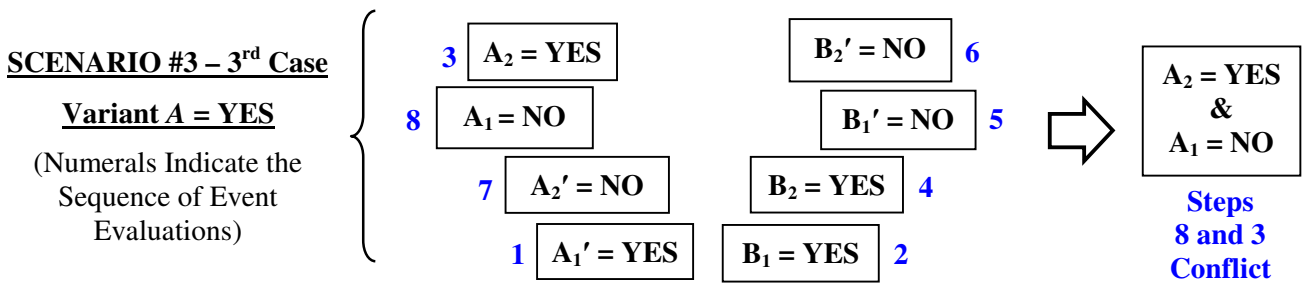

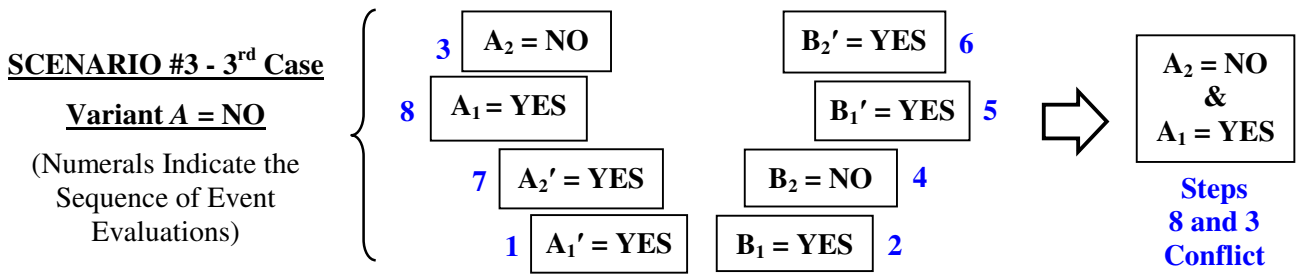

Referring now to variant *B*, assume first that $B_2'$ is YES, so that $A_2'$ is also YES. $A_1$ is then YES, so that $A_2$ is also YES. $B_2$ must then also be YES, and thus $B_1'$ is NO. But $B_2'$ cannot be YES, when the action $B_1'$ is to conduct an entanglement demolishing measurement at $B_2'$. Assume instead then that $B_2'$ is NO, so that $A_2'$ is NO. $A_1$ is then NO, so that $A_2$ is also NO. $B_2$ must then also be NO, and thus $B_1'$ is YES. Again, the assumed result of $B_2'$ conflicts with the action $B_1'$ that determines the type of measurement conducted at $B_2'$:

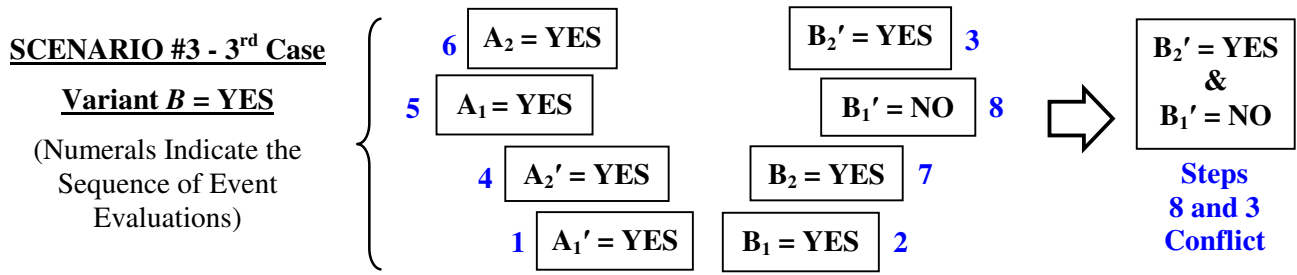

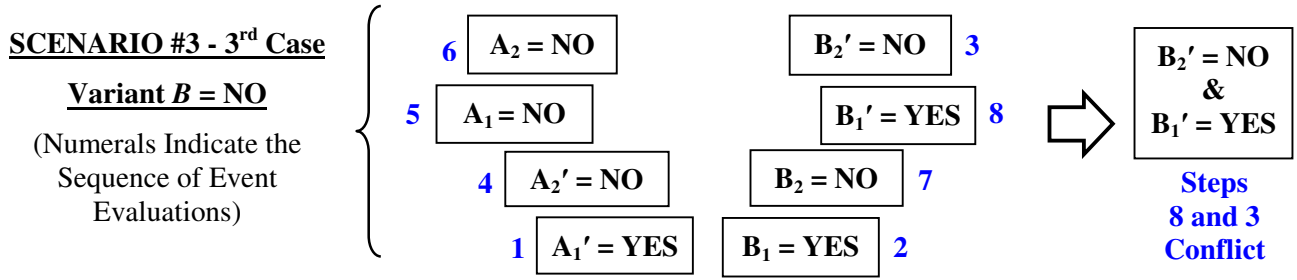

Conducting the third case analysis by assuming what occurs in a later event in one frame before determining what occurs in an earlier event in that frame is justified because all the possible options for the combinations of the first two measurements are included in the four variants. Alternatively, the analysis can be conducted by assuming the result of either first measurement, $A_2'$ or $B_2$, at step **3** of the event evaluations. Conducting these alternate variant analyses finds equivalent ensuing conflicts:

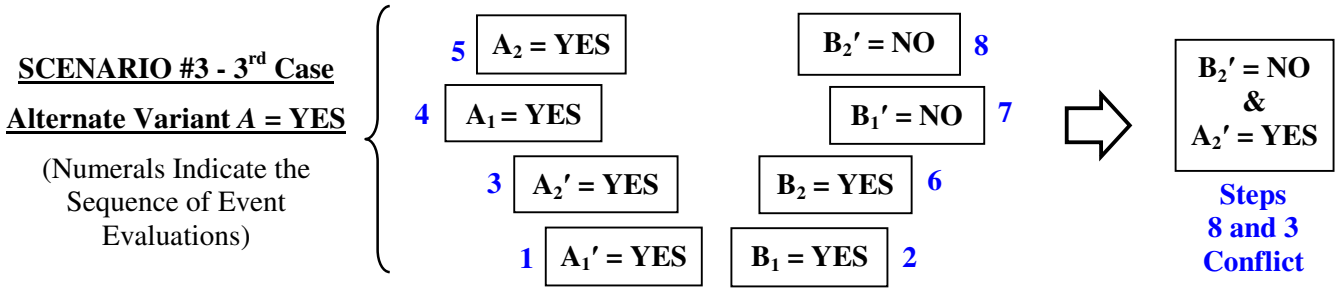

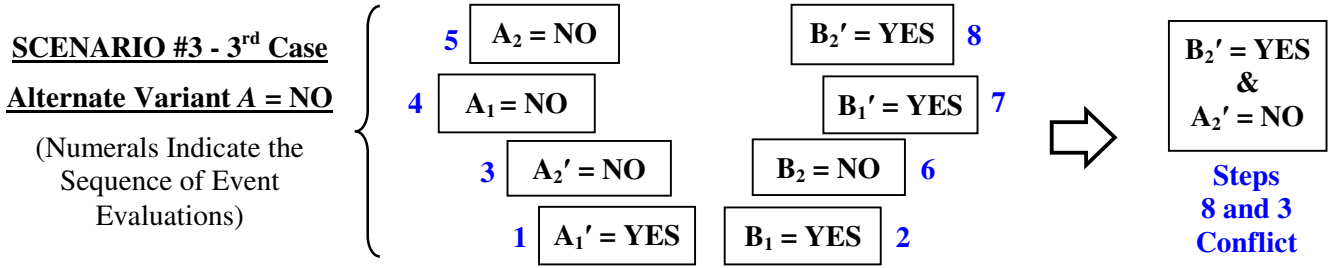

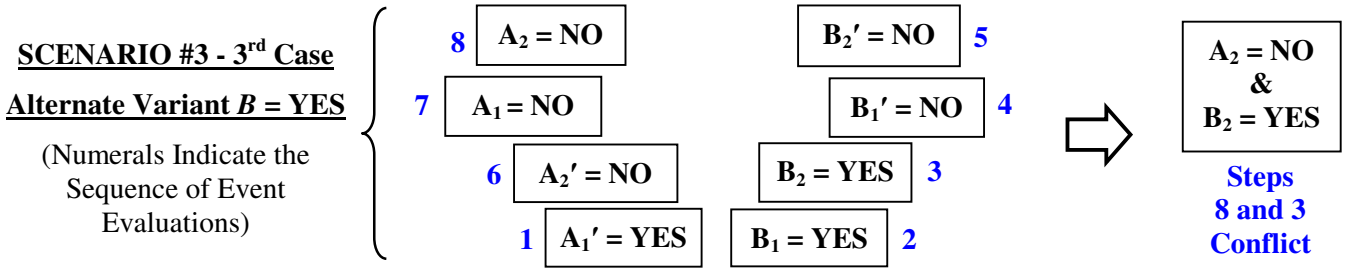

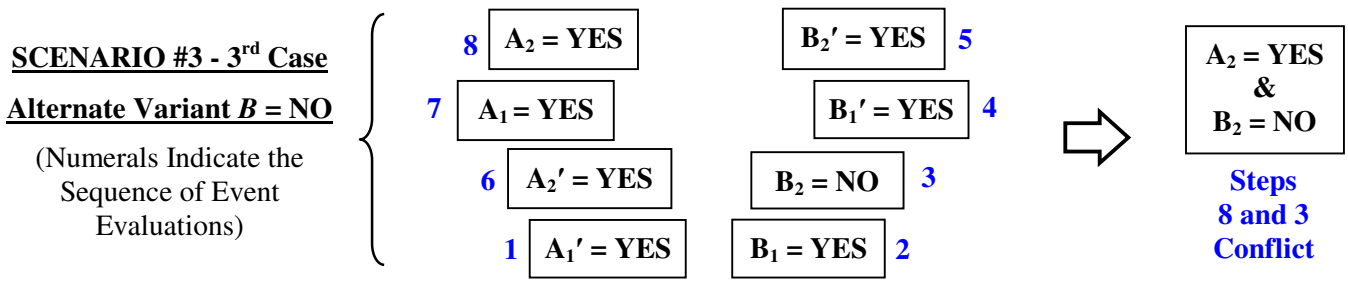

**Summary of Scenario #3 Causal Loop Creation Results**

The third case of the scenario #3 event sequences show that when both observers first action is to conduct an entanglement confirming measurement, subsequent events will conflict. In the first set of variants of the third case, the assumed results of the reference frames' second measurements ($A_2$ or $B_2'$) are incompatible with the prior actions ($A_1$ or $B_1'$, respectively) that determine the assumed measurements' types. For example, in the instance of the variant "*A*=YES" of the first set of variants of the third case, upon ascertaining that the assumed $A_2$ =YES is impossible since $A_1$ =NO, the customary response is to conclude that $A_2$ =NO instead. But, if the events subsequent to the original assumption ($A_2$ =YES) are not discarded, all these events conflict with the consequences of the changed $A_2$ result, and thus, a causal loop is the outcome. When the subsequent events are discarded and the analysis is conducted anew, the event succession is now the variant *A*=NO of the first set of variants of the third case and a causal loop is again the eventual outcome. Causal loops are found to be the outcomes of the rest of the first set of variants of the third case of scenario #3 as well. Likewise, all of the alternate set of variants of the third case of scenario #3 also manifest causal loops explicitly.

<u>*APPENDIX B*</u>

**ENTANGLEMENT CONFIRMING/PRESERVING MEASUREMENT PROCESS**

<u>**Background and Setup**</u>

If a capacity to reliably execute entanglement confirming/preserving measurements can be established, physical implementation of the CLCE should be able to definitively discern the relative accuracies of the Ephemeral TI and the Episodic TI perspectives. The measurement process needs to be capable of confirming the presence of an entanglement (generally by confirming the presence of a superposition of states) without ending that entanglement in the process of confirming its existence. It is not necessary to determine the state of the entangled entity (usually an entanglement demolishing measurement) when confirming an entanglement, but it is necessary to distinguish between the presence and the absence of entanglement. The measurement process should also be switchable to an entanglement demolishing form when necessary, though this alone does not present any significant technical obstacles, and is not addressed in detail here.

The CLCE is primarily envisioned as being conducted through the manipulation and measurement of entangled photons. While it is possible to use electrons, atoms or other entangled entities in place of photons, present technical constraints tend to favor the use of photons, due in no small part to the accelerating rate of experimental accomplishments in quantum optics and quantum computing. As discussed in a review (Tittel, W. and Weihs, G., 2001) of the "state of the art of photonic experiments" that use and explore entanglement, various experiments have been conducted to attempt to close "loopholes" in the debate about the existence of quantum non-locality. Though the question of whether or not a single experiment has managed to close all such loopholes simultaneously and thereby end the debate is still open, all of the loopholes appear to have been closed in at least one or another experiment. This question is not addressed here, wherein the assumption is made that non-local phenomena are real and demonstrated, if not yet perfectly proven. Ideally, the CLCE would accomplish its primary objectives, operate in a completely QND manner wherever desired, and close all loopholes simultaneously. Though this is a worthwhile goal, the host of complications this would comprise is beyond the scope of the present considerations. Accordingly, the present CLCE design is focused on accomplishing its primary aims, while achieving these other goals whenever possible.

Photonic entanglement experiments involve challenges ranging in scale from the substantial to the subtle. In translating the manipulation and measurement of entangled photons from the gedanken to the real world, compromises are inescapable. Issues such as entanglement preparation, source brightness, entanglement purification or distillation, degree of entanglement control and verification, avoidance of unintended decoherences, and measurement results requiring statistical analyses complicate real world efforts. The issue of the fair sampling assumption is also germane, but its resolution seems to be a matter of (presumably transitory) technical constraints. In addition to these complications, achieving the present entanglement confirming/preserving measurement also involves accomplishing a form of QND measurement. Executing an entanglement demolishing measurement is fairly straightforward. When a dual-path interference procedure is employed, merely determining which path an entity takes will end the entanglement. The entanglement confirming measurement process (ECMP), however, is preferably able to both detect the presence of an entangled state without demolishing that entanglement and provide a statistically strong discernment between the presence and the absence of entanglement, all with the same arrangement.

The design objective of the present approach is to detect photons only when the entanglement is absent, so that a failure to detect photons indicates the presence of an entanglement. In this way, when optimally executed, the measurement process will interact minimally with the entangled photons, so that the measurement process is able to preserve the entanglement while confirming its existence. Following up on the work of Ou and Mandel (Ou, Z. Y., Hong, C. K. and Mandel, L., 1987; Ou, Z. Y. and Mandel, L., 1988), a substantial body of research (Kwiat, P., Mattle, K., Weinfurter, H. and Zeilinger, A., 1995; Kwiat, P., Waks, I., White, A. G., Appelbaum, I. and Eberhard, P. H., 1999; Kurtsiefer, C., Oberparleiter, M. and Weinfurter, H., 2001) experimenting with photons in entangled orthogonal states of polarization from type II spontaneous parametric down conversion has arisen. The use and manipulation of polarization entangled photons from this type of source, as well as their limitations and remaining difficulties are now well studied. The entangled state to be used will be a superposition of vertically and horizontally polarized eigenstates $|\mathbf{v}\rangle_{S\,(L)}$ and $|\mathbf{h}\rangle_{L\,(S)}$ of first and second photons (where the subscript L indicates traveling along the longer elapsed travel time, i.e. delay coil, fiber and the subscript S indicates traveling along the shorter elapsed travel time, i.e. direct, fiber). It has been shown (Kwiat, P., Mattle, K., Weinfurter, H. and Zeilinger, A., 1995; Weihs, G., Jennewein, T., Simon, C., Weinfurter, H. and Zeilinger, A., 1998; Michler, M., Weinfurter, H., and Żukowsk, M., 2000) that with the use of birefringent crystals, quarter and half wave plates, and other phase shifters that any of the four EPR-Bell states can be produced. For the present experiment, the entangled state used will be the Bell state:

$$\Psi_+ = 1/\sqrt{2}\ \left[|\mathbf{v}\rangle_L|\mathbf{h}\rangle_S + |\mathbf{h}\rangle_L|\mathbf{v}\rangle_S\right] \tag{1}$$

Purification and production efficiency of this state are important issues both for the statistical strength of the results, as well as for the satisfaction of the fair sampling assumption. But these technical challenges are not significantly different for this experiment than they are for the majority of experiments with entangled photons, and various approaches (Kwiat, P., Mattle, K., Weinfurter, H. and Zeilinger, A., 1995; Kwiat, P., Waks, I., White, A. G., Appelbaum, I. and Eberhard, P. H., 1999; Kurtsiefer, C., Oberparleiter, M. and Weinfurter, H., 2001; Zhao, Z., Pan, J.-W. and Zhan, M. S., 2001) to satisfying these questions are being employed with increasingly greater success.

The photons should be emitted in a series of entangled pairs that are sufficiently spaced in time from the immediately preceding and following photon pairs so that each photon's interaction with an observer's measurement apparatus is separate from the preceding and following photons' interactions. The two sources' emissions of photon pairs are to be sufficiently synchronized so that the pairs are emitted within a time interval that is both much shorter than the time intervals between any of the events in either observer's reference frame, and is also much shorter than the difference in elapsed travel time along the direct versus the delayed optical fibers. The time interval between successive photon pair emissions should be substantially more than the greatest aggregate time required for a photon to traverse a delayed fiber plus the time required to execute all of the observer's actions and measurements on that photon. The measurement process begins prior to the photons entering the portion of the apparatus depicted in Figure 6 (which is en masse termed the "entanglement detector").

Figure 6 schematically depicts an apparatus for conducting the ECMP. Initially, $\Psi_+$ photons are directed (by optical fibers) through a birefringent crystal which is aligned so that it splits the vertical and horizontal polarization components as described previously (Michler, M., Weinfurter, H., and Żukowsk, M., 2000). Since the photons are in a superposition of these polarization states, they take both paths equivalently as long as the paths are not monitored to detect the photon's passage. The ECMP operates similarly for photons arriving by either the shorter or the longer path. The ECMP's operation is described here for photons which arrive by the longer path (the description for photons which arrive by the shorter path is comparable). After passing through the birefringent crystal the horizontally polarized component is directed along optical fiber 1 and the vertically polarized component is directed along optical fiber 2. The state of the photon is then:

$$\Psi_+' = 1/\sqrt{2} \; \left[ \left| v \right\rangle_L \left| h \right\rangle_S \left| 2 \right\rangle_L + \left| h \right\rangle_L \left| v \right\rangle_S \left| 1 \right\rangle_L \right] \tag{2}$$

Because orthogonal polarizations will not interfere with each other, it is necessary to pass one of the two paths through a phase shifter to align the polarizations of the photon components taking one path with the polarization of those taking the other path. It has been shown (Kwiat, P., Mattle, K., Weinfurter, H. and Zeilinger, A., 1995) that "…a half wave plate in one path can be used to change horizontal polarization to vertical and vice versa." Accordingly, optical fiber 1 passes the photon through a half wave plate, so that its state is then:

$$\Psi_+'' = 1/\sqrt{2} \; \left[ \left| v \right\rangle_L \left| h \right\rangle_S \left| 2 \right\rangle_L + \left| v \right\rangle_L \left| h \right\rangle_S \left| 1 \right\rangle_L \right] \tag{3}$$

The ensuing photon state $\Psi_+''$ is capable of exhibiting "self-interference" when the $\left| 1 \right\rangle_L$ and $\left| 2 \right\rangle_L$ components are subsequently recombined.

Among the constraints on the utilizable photon wavelengths (Tittel, W. and Weihs, G., 2001) are optical fibers' technical limitations, such as absorption minima confined within tight wavelength ranges, as well as the restricted availability of high efficiency and low noise single photon detectors/counters. Because the wavelengths of the entangled photons will be known, dynamic phase differences between the two paths can be selectively controlled by an auxiliary phase shifter (not shown in Figure 6) and/or careful management of the path lengths along the fibers. The phase shifter could also be used to account for the $\pi/2$ phase shift between the reflected and the transmitted components of the photons (Degiorgio, V., 1980).

### Differentiation Between Entanglement Presence/Absence

The accuracy of the ECMP will depend in part on the degree of precision with which the entanglement detectors depicted in Figure 6 are able to predictably and precisely produce interference between the $\left|2\right\rangle_L$ and the $\left|1\right\rangle_L$ components of $\Psi_+''$. For a given photon wavelength $\lambda$, the separation **a** between the emitting ends of fibers **1** & **2** and the distance **L** between the ends of the fibers and the "detection" screen location can be varied to selectively control the positioning of destructive interference bands at the "detection" screen. The fibers are preferably maintained in straight, parallel orientations at right angles to the "detection" screen. Positions **1** & **2** on the "detection" screen are directly in line with the continuations of the paths of the fibers **1** & **2**, respectively. The objective of the ECMP is to accomplish the greatest degree of destructive interference at positions **1** & **2**. This will occur when the distance from the end of fiber **1** to position **2** on the "detection" screen is ½ $\lambda$ longer than the distance from the end of fiber **2** to position **2**, and vice versa.

Optimally, the presence (vs. absence) of entanglement of the photons entering the entanglement detectors correlates with maximum variation in the probability of photons being detected at positions **1** and **2**. When a photon is entangled in the prescribed state $\Psi_+''$ locating the photon detectors' apertures at the positions of greatest destructive interference will minimize their probability of receiving a photon.

Consequently, the detectors' will register the lowest combined counting rate of entangled photons when their apertures are at positions **1** and **2**.

As discussed in the literature (Zeilinger, A., 1999), the potential of being able, even in principle, to later execute a measurement that can recover an entangled photons' path information can eradicate interference effects. One manner (Zeilinger, A., 1999) of eliminating this issue is to use a Heisenberg detector and lens arrangement to project these photons into momentum eigenstates which cannot reveal any position information, and hence cannot reveal any path information either. At first glance, it may seem necessary to ensure that the photons' which-path entanglement (equivalent to the photon being in an entangled state when received by the ECMP) is not recoverable after the photons have passed the "detection" screen in the ECMP. However, this information does not appear to be recoverable in the presently described experimental setup. At any rate, a different technique might be central to the delayed choice protocol functioning as desired. Ensuring that the entangled, and hopefully "undisturbed", photons remain intact after they pass the detection screen, at least until the delayed measurements are actually executed, may prove more fruitful. Directing the photons into an optical fiber storage ring or double mirror assembly might maintain the entangled photons' sufficiently long enough to accomplish the delayed choice measurements.

There are two intended ways of ending a photon's entanglement in the CLCE. The first way is by undergoing a polarization determining measurement itself, and the second is through a polarization determining measurement of its entangled counterpart photon. In both cases, the polarization determination is conducted by activating the selectable photon counters of the ECMP. Preferably, the counters are QND in operation, though this is not essential and the technical issues involved are not only unresolved but may present complications for maintaining the fidelity of the entanglements being measured. When the counters are active, they will determine which path the photon takes after the birefringent crystal and thereby project the photon into one or the other polarization eigenstate. (When the counters are not QND, the registration by either counter of a photon in one of the fibers would then represent the measurement result "entanglement missing", rather than differences in detection events registered by the photon detectors. The counting of a photon along either of the fibers is a definitive indication that the entanglement is thereafter ended, regardless of whether or not the entanglement was intact prior to that counting.) An unentangled photon is emitted from only one of the optical fibers, and whether or not it is known which fiber it was emitted from, it will not exhibit interference effects. Photons that are unentangled cross the "detection screen" in Gaussian distributions centered at positions **1** & **2**. Consequently, locating the detectors' apertures at positions **1** and **2** will register the greatest combined counting rate when the photons are unentangled.

The strategy herein employed attempts to maximize the difference in the counting rates at the two photon detectors between entangled and unentangled photons. Besides the separation **a** and the distance **L**, the width **d** of the receiving aperture of the photon counters is also significant. Due to technical constraints such as experimental imprecision, interference effects cannot be perfectly evaluated. Even if these constraints were surmountable, the theoretical width of complete destructive interference is vanishingly small. As a result, it is not possible, even in theory, to dispose a real counter of non-zero aperture width **d** entirely within the area of complete destructive interference. Additionally, while a minimized aperture width **d** will decrease spurious photon counts when the photons are entangled, it will also reduce the rate of counting of photons when the entanglement is absent. The greatest statistical strength of differentiation between the presence and absence of entanglement will require experimentation with variations in the width **d**, for a given λ, **a**, and **L**.

While the strategy described above is not perfectly QND, it does approach at least a degree of "Quantum Non-Disturbance" that is an appealing cousin to an actual QND measurement. Since the objective is to interact with as few photons as possible when confirming and preserving entanglement, it is preferable for the counters to intersect with the emitted photons only at their respective apertures. Accordingly, restricting the entanglement detector to a narrow 2-dimensional field may enable the remainder of the counter apparatuses to be disposed outside of the plane of Figure 6 and thereby evade unwanted interactions. (Though the counters in Figure 6 are shown as extending past the "detection screen" away from the fibers, this is not the best layout, since the positions of maximum destructive interference vary with increasing distance, and at distances greater than **L** any placement of the counters in the field of the entanglement detector will result in unwanted interactions with entangled photons.)

The above approach produces a statistical determination of the presence/absence of entanglement, with potentially large degrees of certainty. The statistical approach is capable of being definitive, since large numbers of independent photons subject to the same conditions, rules, and measurement actions are measurable as a group. At minimum, the arrival time of the group as a whole should be brief enough to enable the entire group of short fiber photons from one source to interact with an observer's first measurement apparatus before that observer's second measurement receives any of the group of long fiber photons from the other source. The interim between group interactions may also need to be sufficiently protracted to further allow execution of the first measurement, communication of the first measurement result, application of the rule applicable to the first measurement result, and implementation of the choice of the type of the second measurement. A delayed choice protocol could prove felicitous for carrying out these steps within the allotted interval. The measurements and actions ought to still be executed in the prescribed relative orders, but the time available for the completion of each as well as the time periods available between the completion of one measurement or action and the execution of the next might be more flexible when conducted in delayed choice manners.

It is, of course, preferable to positively detect the presence/absence of entanglement photon by photon. A revision of the above strategy may accomplish this aim, though the technical obstacles are likely more rigorous. In such a single-photon entanglement detection strategy, the fibers do not terminate in photon emitting outlets separated by a distance **a**. The fibers' ending sections are instead aimed in trajectories that cross at a shallow angle θ. The fibers would terminate just before they intersect, and they could typically emit their photons into a minimally dispersive media (such as an evacuated chamber). The path lengths along each of the fibers can be controlled so that, when the photon is in an entangled state, the emitted photon components destructively interfere where the trajectories cross. Realizing a reliable means of distinguishing the presence of a photon (indicating the photon is not entangled) from the absence of a photon (indicating the photon is entangled) at the paths' intersection is among the chief technical challenges to single-photon entanglement detection.

A first prospective distinguishing method is to utilize a birefringent crystal placed with its optical axis aligned so that it will redirect a photon in a polarization eigenstate at a known angle. The redirected photon is then directed along an uptake optical fiber to a detector. It is necessary to align the birefringent crystal so that it redirects both polarization eigenstates away from the continuation of either's trajectory when emitted by the ending sections. Each of the redirected polarization eigenstates are directed along uptake optical fibers to detectors. The aim is for an entangled photon to be destructively interfering where the paths cross at the birefringent crystal so that it will not interact with, and not be redirected by, the crystal. Birefringent crystals split the propagation directions of photons with different polarizations because the propagation speed through the crystal is different for different polarization orientations. In order to accomplish a significant enough redirection of a photon, the crystal must have a sufficient width for the photon passing through it to experience an adequate change in velocity. Given that the wavelength of the photons to be used are typically of the order of $10^{-7}$ meters, it seems clear that the region where the paths intersect will have to extend for a large number of wavelengths for the crystal to effect any appreciable change in direction of a photon. Whether it is possible for the region of destructive interference to be large enough is an open question, though at minimum the crossing angle $\theta$ will likely need to be very small. Besides the two uptake fibers that direct reflected photons to detectors, two other fibers will direct unreflected, entangled photons from their original paths to momentum measurements that will preserve their position (and hence polarization) entanglements (Zeilinger, A., 1999).

A second prospective distinguishing method is for the paths to intersect at an optical Kerr media. In theory, the phase of a probe entity such as a photon or an atom within the optical Kerr media will be altered when a signal photon passes through the optical Kerr media concurrently. A signal photon will be present (i.e. not be destructively interfering) when crossing the optical Kerr media, and thereby be available to induce the probe entity's phase alteration, only when the photon undergoing the ECMP is not entangled. In practice, reliably detecting an appreciable probe entity phase alteration due to the presence of a solitary signal photon is problematic.

A form of QND measurement that would utilize a Mach-Zender interferometer could also prove useful. In this approach, the Mach-Zender interferometer is arranged so that probe entities traversing its arms destructively interfere upon recombination, absent outside influences. A detector is arranged so as to register any probe entities when the destructive interference does not occur. It may be possible to adapt this approach so that at least one arm of the Mach-Zender interferometer is comprised of an optical Kerr media. The passage of the signal photon through this optical Kerr media arm might provide sufficient phase disturbance to compromise the destructive interference of the probe entities and enable their registration by the detector.

An alternative prospective manner of conducting a single-particle ECMP is to use spin polarized particles such as electrons in a Stern-Gerlach interferometer. The use of massive particles for this form of interferometry is well-known, and their utilization in delayed choice experiments has also been demonstrated (Lawson-Daku, B. J., Asimov, R., Gorceix, O., Miniatura, Ch., Robert, J. and Baudon, J., 1996). For the present purposes, analogues to the components of the CLCE which control photons are available for electrons also. One difference is the use of a different Bell state:

$$\Phi_+ = 1/\sqrt{2} \left[ \, |\uparrow\rangle_L |\uparrow\rangle_S + |\downarrow\rangle_L |\downarrow\rangle_S \, \right] \qquad (4)$$

The paths of the entangled components of this state can be directed to follow curved paths by electromagnetic fields. These paths can be brought into greater lengths of coincidence than is available with the intersecting photon paths. In this expanded area of path intersection, where the ECMP will produce destructive interference of the $\Phi_+$ entangled components, a localized electric field can direct an unentangled electron to a detector.

The feasibility of the strategies outlined will be subject to evaluation (and likely alteration) by experimentalists, and their successful completion will require no small amount of skill and determination. Although the technical obstacles to realizing the ECMP are not insignificant, and may even be presently beyond attainment for the single particle strategies, the experimental results should be both highly significant and convincing.

## REFERENCES

Aspect, A., *Nature* **398**, 189 (1999).

Bell, J. S., *Physics* **1**, 195 (1964).

Bohm, D., *Physical Review* **85**, 166 and 180 (1952).

Bohm, D. and Aharonov, Y., *Physical Review* **108**, 1070 (1957).

Bohr, N., *Physical Review* **48**, 696 (1935);  N. Bohr, *Nature* **136**, 1025 (1935).

Brezger, B., Hackermüller, L., Uttenthaler, S., Petschinka, J., Arndt, M.  and Zeilinger, A., *Physical Review Letters* **88**, 100404 (2002).

Cramer, J. G., *Reviews of Modern Physics* **58**, 647 (July 1986).

Degiorgio, V.,  *American Journal of Physics* **48**, 81 (1980).

DeWitt, B. S., and Graham, N., (eds), *The Many-Worlds Interpretation of Quantum Mechanics* (Princeton: Princeton University Press, 1973).

Einstein, A., Podolsky, B., and Rosen, N., *Physical Review* **47,** 777 (1935).

Everett, H., *Reviews of Modern Physics* **29**, 454 (1957).

Franson, J. D., *Physical Review D* **31**, 2529 (1985); Interestingly, the author has become aware only in the very final stage of completing this paper that on p. 2531 Franson offers that "…an alternative interpretation would be to abandon the idea of there being discrete events, and to view the entire situation, including the measurement apparatus, as a *continuous process.*" (emphasis added)  This is the first and only inclination offered by any, insofar as the author is aware, to even begin to consider the new perspective presented in the present paper.  Apparently neither Franson, nor any others, considered the possibility of the present hypothesis.

Freedman, S. J.  and Clauser, J. F., *Physical Review Letters* **28**, 938 (1972).

Friedman, J. R., Patel, V., Chen, W., Tolpygo, S. K. and Lukens, J. E., *Nature* **406**, 43 (2000).

Grangier, P., *Nature* **409**, 774 (2001).

Hardy, L., *Physical Review Letters* **68**, 2981 (1992).

Kim, Y.-H., Yu, R., Kulik, S. P., Shih, Y. and Scully, M. O., *Physical Review Letters* 84 (2000).

Kurtsiefer, C., Oberparleiter, M. and Weinfurter, H., *Physical Review A* **64**, 023802-1 (2001).

Kwiat, P., Mattle, K., Weinfurter, H. and Zeilinger, A., *Physical Review Letters* **75**, 4337 (1995).

Kwiat, P., Waks, I., White, A. G., Appelbaum, I. and Eberhard, P. H., *Physical Review A* **60**, R773 (1999).

Lawson-Daku, B. J., Asimov, R., Gorceix, O., Miniatura, Ch., Robert, J. and Baudon, J., *Physical Review A*, **54**, 5042 (1996).

Mermin, N. D., *American Journal of Physics* **66**, 753 (1998). (Mermin, N. D., 1998)

Michler, M., Weinfurter, H., and Żukowsk, M., *Physical Review Letters* **84**, 5457 (2000).

Myatt, C. J., King, B. E., Turchette, Q. A., Sackett, C. A., Kielpinski, D., Itano, W. M.,

Monroe, C. and Wineland, D. J., *Nature* **403**, 269 - 273 (2000).

Ou, Z. Y., Hong, C. K. and Mandel, L., *Physics Letters A* **122**, 11 (1987).

Ou, Z. Y. and Mandel, L., *Physical Review Letters* **61**, 50 (1988).

Percival, I. C., *Physics Letters A* **244**, 495 (1998); quant-ph/9803044.

Percival, I. C., *Proceedings of the Royal Society of London* **A 456**, 25 (2000); quant-ph/9811089.

Percival, I. C., quant-ph/9906005.

Schrödinger, E., *Die Naturwissenschaften* **23**, 807 (1935).

Tittel, W., Brendel, J., Zbinden, H. and Gisin, N., *Physical Review Letters* **81**, 3563 (1998).

Tittel, W. and Weihs, G., *Quantum Information and Computation* **1**, 2 (2001), 3-56; and references therein.

Visser, M., gr-qc/0204022, *The Quantum Physics of Chronology Protection*, from *The Future of Theoretical Physics and Cosmology*, Celebrating Stephen Hawking's 60th Birthday (2003).

von Neumann, J., *Mathematische Grundlagen der Quantenmechanik*, (Berlin: Springer-Verlag 1932); *Mathematical Foundations of Quantum Mechanics* English Trans.-Robert T. Geyer (Princeton: Princeton University Press, 1955).

Weihs, G., Jennewein, T., Simon, C., Weinfurter, H. and Zeilinger, A., *Physical Review Letters* **81**, 5039 (1998).

Wheeler, J. A., in *Mathematical Foundation of Quantum Theory*, edited by A. R. Marlow (Academic, New York, 1978).

Zbinden, H., Brendel, J., Gisin, N.  and Titel, W., *Physical Review A* **63**, 022111-1 (2001).

Zeilinger, A., *Reviews of Modern Physics* **71**, S288 (1999).

Zhao, Z., Pan, J.-W. and Zhan, M. S., *Physical Review A* **64**, 014301-1 (2001).

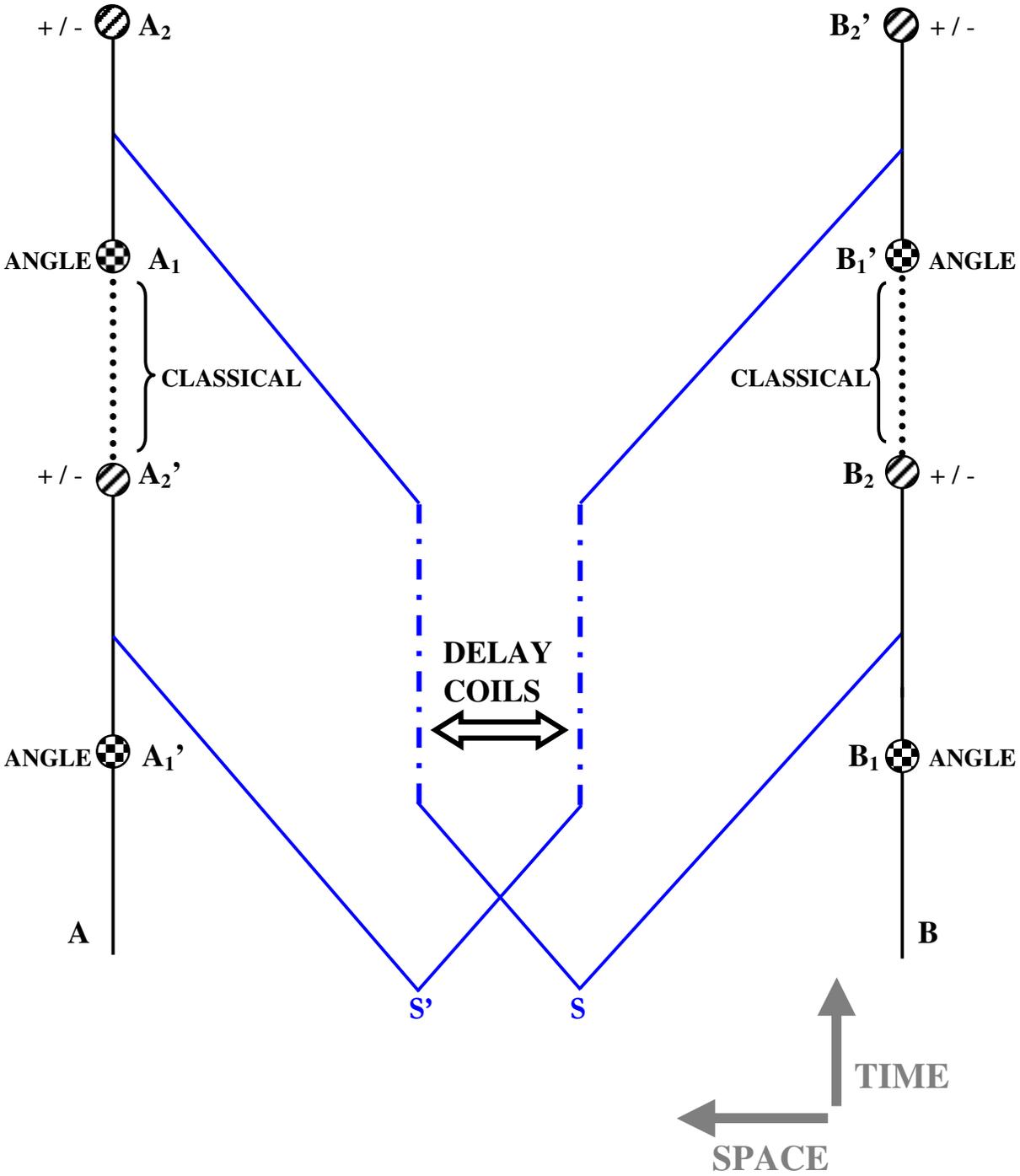

**Figure 1.** Percival's Figure 3 "Spacetime diagram of the double Bell experiment" (adapted)
    In this figure, the spatial dimension is shown as compacted, whereas in Percival (2000), all of the events in either reference frame are spacelike separated from all the relevant events in the other reference frame. Experimental settings are made at $A_1$, $A_1'$ and $B_1$, $B_1'$; measurements occur at $A_2$, $A_2'$ and at $B_2$, $B_2'$. The result of the measurement at $A_2'$ (+ or -) is classically communicated to affect the setting at $A_1$ which is used to make measurement $A_2$; and the result of the measurement at $B_2'$ (+ or -) is classically communicated to affect the setting at $B_1$ which is used to make measurement $B_2$.

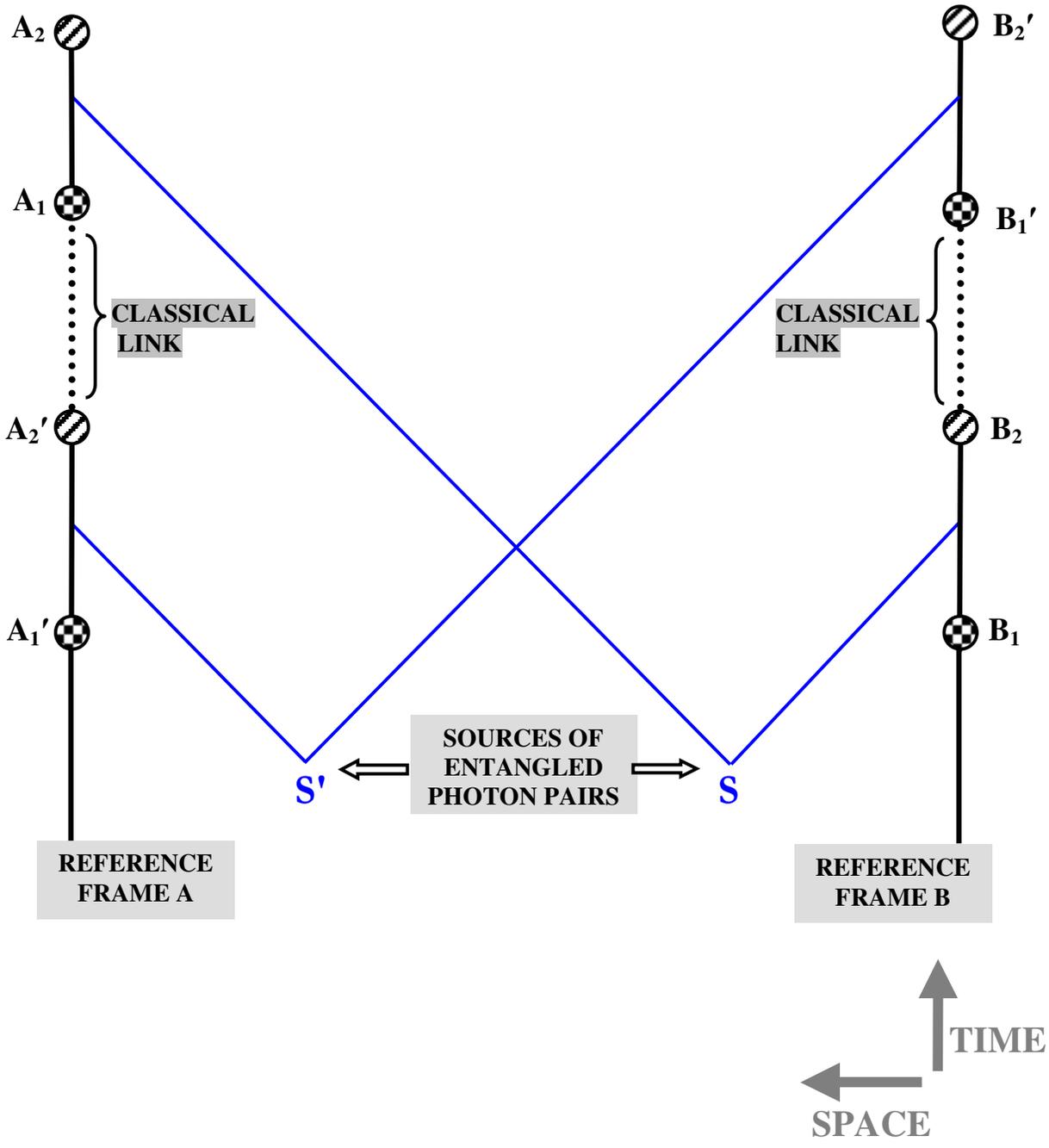

**LEGEND**

$A/B_1^{(')}$  ⟺ Decision on Type of Measurement

$A/B_2^{(')}$  ⟺ Interaction of Photon(s) with Measurement Apparatus

**Figure 2.** Schematic Set-Up of the Causal Loop Creation Experiment

**FIGURE 3**

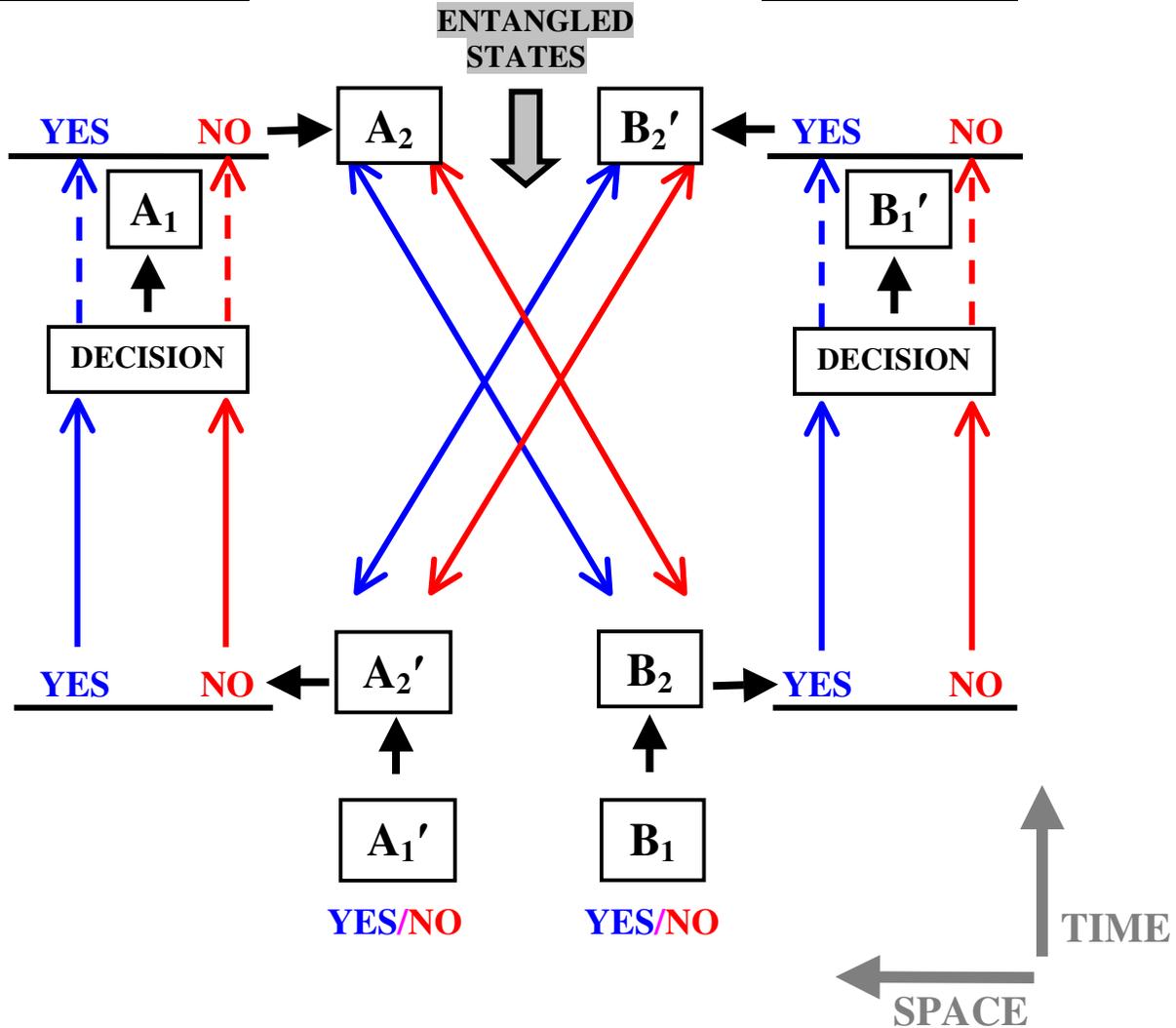

# OBSERVER A

*Decision Rule:*
*ACTION SAME AS*
*OBSERVATION*

# OBSERVER B

*Decision Rule:*
*ACTION SAME AS*
*OBSERVATION*

ENTANGLED
STATES

YES   NO   A$_2$   B$_2'$   YES   NO

A$_1$   B$_1'$

DECISION   DECISION

YES   NO   A$_2'$   B$_2$   YES   NO

A$_1'$   B$_1$

YES/NO   YES/NO

TIME

SPACE

---

**LEGEND FOR FIGURES 3-5**

**YES** ≡ Entanglement Confirming (observation) / Entanglement Preserving (action)

**NO** ≡ Entanglement Missing (observation) / Entanglement Demolishing (action)

• Subscript "1" denotes an action   • Subscript "2" denotes a observation

---

**Figure 3.**   **Scenario #1: Causal Loop Non-Creation**
The action/observation diagram of the Causal Loop Creation Experiment for scenario #1. When both observers' actions correspond with their observations, a causal loop is not created, in the ideal case without unintended decoherence.

**FIGURE 4**

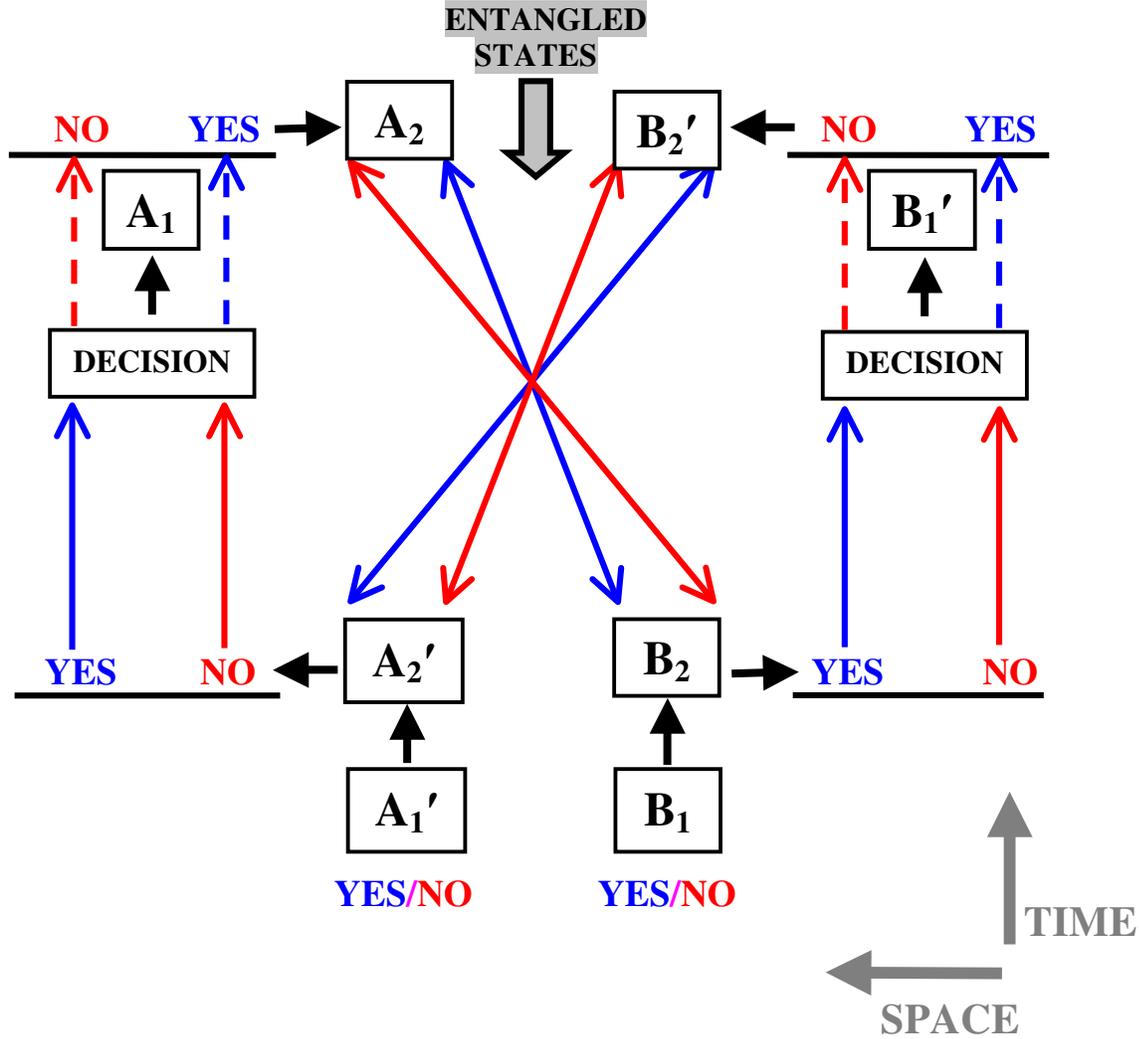

# OBSERVER A

*Decision Rule:*
*ACTION OPPOSITE*
*OF OBSERVATION*

# OBSERVER B

*Decision Rule:*
*ACTION OPPOSITE*
*OF OBSERVATION*

ENTANGLED
STATES

NO  YES      **A₂**      **B₂′**    NO  YES

**A₁**                              **B₁′**

DECISION                        DECISION

YES  NO      **A₂′**      **B₂**    YES  NO

**A₁′**      **B₁**

YES/NO      YES/NO

TIME

SPACE

---

**Figure 4.    Scenario #2: Causal Loop Non-Creation**
The action/observation diagram of the Causal Loop Creation Experiment for scenario #2.
When both observers' actions are opposite of their observations, a causal loop is not created, in
the ideal case without unintended decoherence.

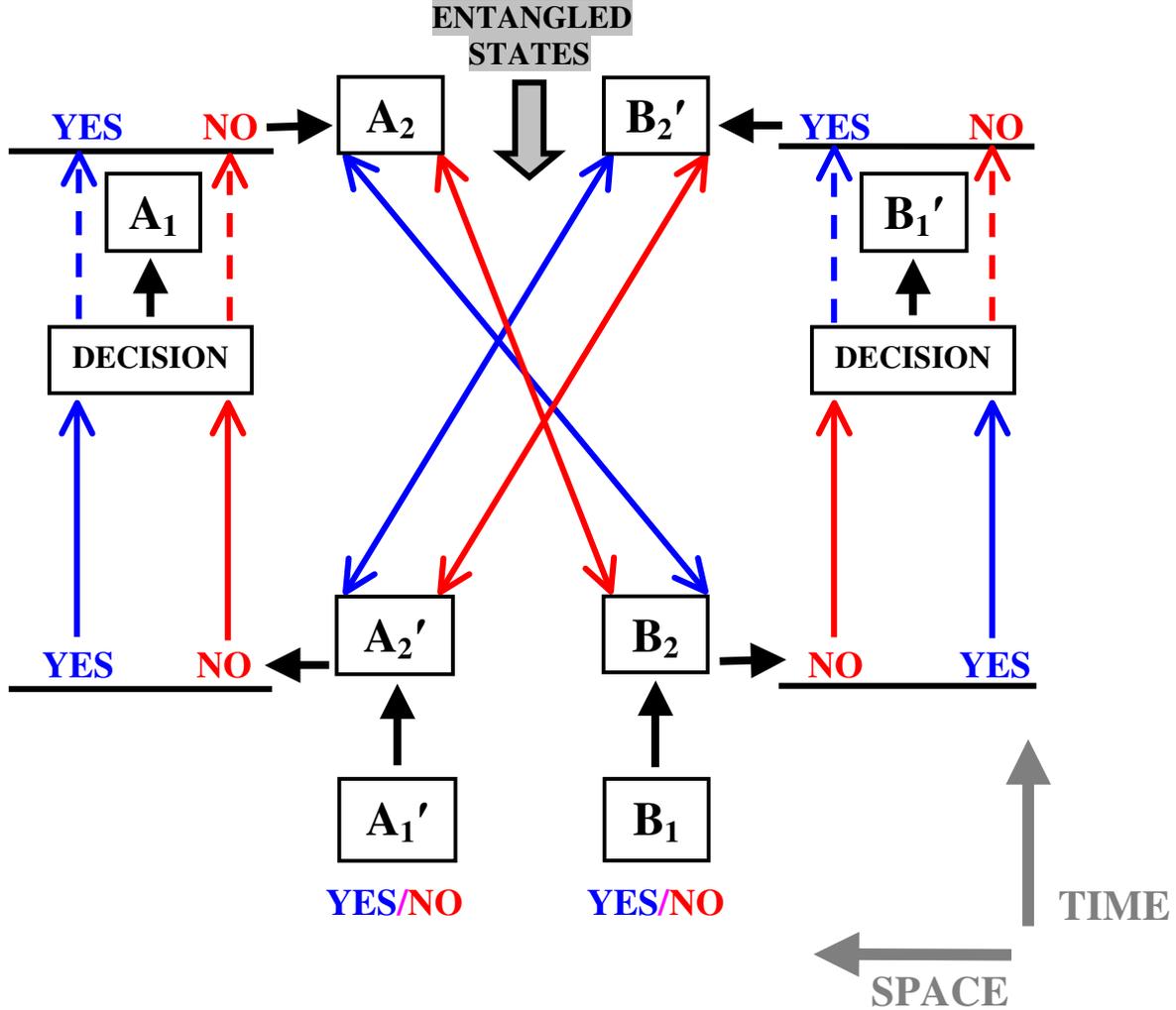

**FIGURE 5.** Scenario #3: Causal Loop Creation

The action/observation diagram of the Causal Loop Creation Experiment for scenario #3. When one observer's actions are opposite of that observer's observations, and the other observer's actions correspond with that observer's observations, a causal loop is created.

**FIGURE 6**

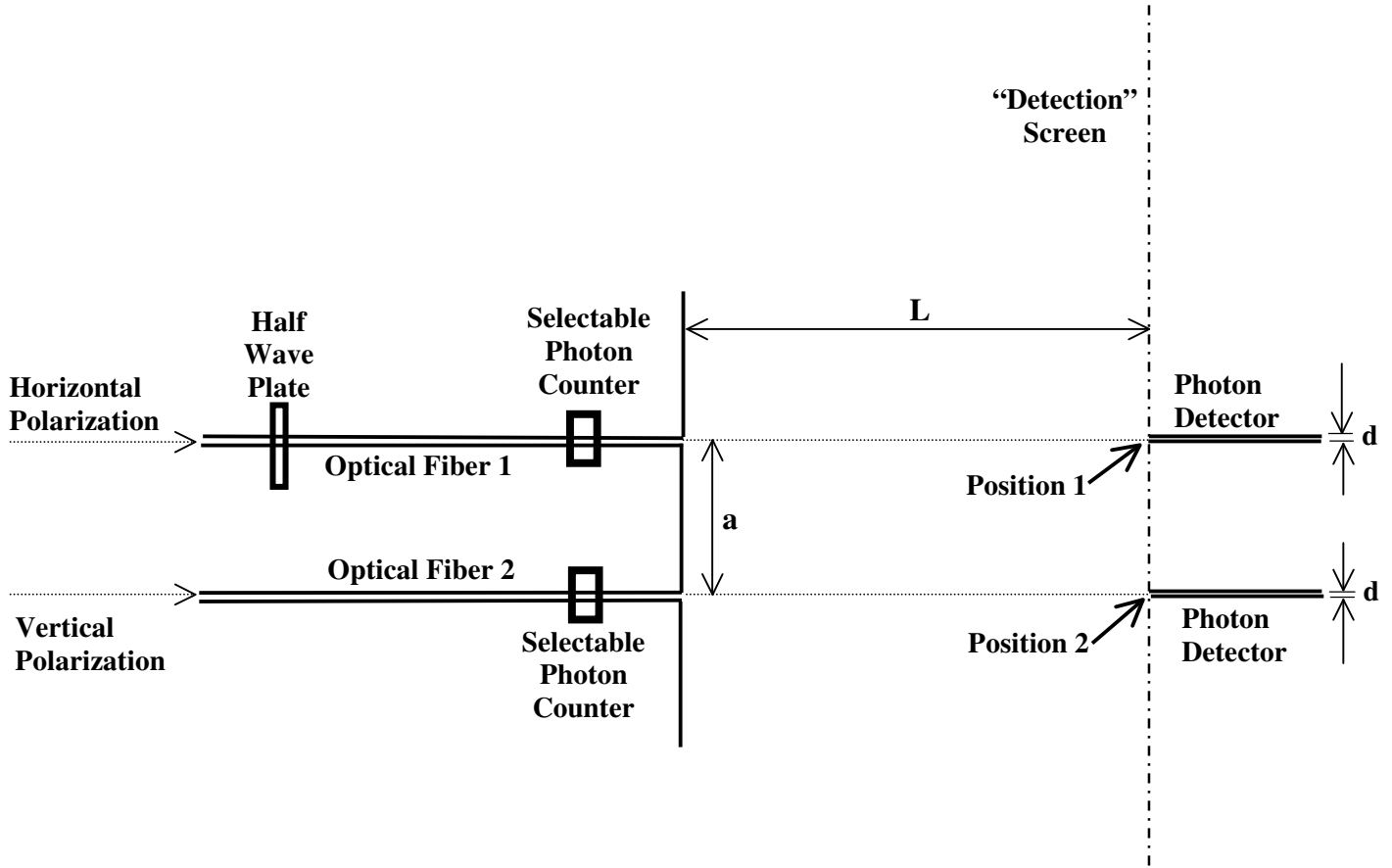

**Fig. 6 Legend**

**a** ≡ Separation between optical fibers

**L** ≡ Spacing between optical fibers and photon detector apertures

**d** ≡ Width of photon detector apertures

<u>**Figure 6.**</u>    **Entanglement Detection Schematic Diagram**
The Entanglement Detection schematic depicts one arrangement for constructing an entanglement confirming apparatus with a selective choice between an entanglement demolishing measurement (by detecting the polarization along the optical fibers) and a minimally demolishing entanglement confirming measurement (in the limit of optimal experimental control and a small photon counter width d).